\documentclass[aps,pra,twocolumn,superscriptaddress,10pt]{revtex4-2}
\usepackage{multirow,amsthm,amssymb,amsbsy,amsmath,epsfig,epstopdf,bm,float}

\usepackage{graphicx}
\usepackage{booktabs}
\usepackage{verbatim}
\usepackage{dcolumn}
\usepackage{color}
\usepackage[dvipsnames]{xcolor}
\usepackage[normalem]{ulem}
\usepackage{soul}
\usepackage{braket}
\usepackage[export]{adjustbox}

\begin{document}

\title{Dynamical phase transitions in quantum reservoir computing}

\author{Rodrigo Mart\'inez-Pe\~na}
\email{rmartinez@ifisc.uib-csic.es }
\affiliation{Instituto de F\'{i}sica Interdisciplinar y Sistemas Complejos (IFISC, UIB-CSIC), Campus Universitat de les Illes Balears E-07122, Palma de Mallorca, Spain}

\author{Gian Luca Giorgi}
\affiliation{Instituto de F\'{i}sica Interdisciplinar y Sistemas Complejos (IFISC, UIB-CSIC), Campus Universitat de les Illes Balears E-07122, Palma de Mallorca, Spain}

\author{Johannes Nokkala}
\affiliation{Instituto de F\'{i}sica Interdisciplinar y Sistemas Complejos (IFISC, UIB-CSIC), Campus Universitat de les Illes Balears E-07122, Palma de Mallorca, Spain}

\author{Miguel C. Soriano}
\affiliation{Instituto de F\'{i}sica Interdisciplinar y Sistemas Complejos (IFISC, UIB-CSIC), Campus Universitat de les Illes Balears E-07122, Palma de Mallorca, Spain}

\author{Roberta Zambrini}
\email{roberta@ifisc.uib-csic.es }
\affiliation{Instituto de F\'{i}sica Interdisciplinar y Sistemas Complejos (IFISC, UIB-CSIC), Campus Universitat de les Illes Balears E-07122, Palma de Mallorca, Spain}

\begin{abstract} 
Closed quantum systems exhibit different dynamical regimes, like Many-Body Localization or thermalization, which determine the mechanisms of spread and processing of information. Here we address the impact of these dynamical phases in quantum reservoir computing, 
an unconventional computing paradigm recently extended into the quantum regime that exploits dynamical systems to solve nonlinear and temporal tasks. We establish that the thermal phase is naturally adapted to the requirements of quantum reservoir computing and report an increased performance at the thermalization transition for the studied tasks. Uncovering the underlying physical mechanisms behind optimal information processing capabilities of spin networks is essential for future experimental implementations and provides a new perspective on dynamical phases.

\end{abstract}
\maketitle

\textit{-- Introduction.} Unconventional computing is an interdisciplinary branch of science that aims to uncover new computing and information processing mechanisms  in physical, chemical and biological systems \cite{jaeger2020exploring}. In this field, the challenge is to develop a device theory guaranteeing that a given system, used as an analog computer, is able to accomplish a computational task.   When it comes to solving temporal tasks, a natural  ``computer" is represented by a system exhibiting   rich dynamical properties. An example of  such approach can be found in reservoir computing (RC), an unconventional  framework belonging to the broad family of  machine learning, derived from recurrent neural networks \cite{jaeger2001echo,maass2002real,lukovsevivcius2009reservoir} but with
the major advantage  of low training cost and fast learning. RC is also especially suited for hardware implementations \cite{appeltant2011information,van2017advances,torrejon2017neuromorphic,tanaka2019recent}. 

For big-data processing, an exceptional playground where a rich dynamics can be exploited is certainly provided by quantum systems,  whose  exponentially large number of  degrees of freedom pushes them towards computational limits that are not achievable by classical systems \cite{Arute2019}. This is the potential envisaged in quantum reservoir computing (QRC) \cite{fujii2017harnessing,mujal2021opportunities}, as recently explored in spin-based implementations \cite{fujii2017harnessing,nakajima2019boosting,chen2019learning,tran2020higher,martinez2020information}, continuous-variable bosonic systems \cite{ghosh2019quantum,ghosh2019neuromorphic,nokkala2020gaussian,govia2020quantum} and fermionic setups \cite{ghosh2020universal}. Efforts to demonstrate proof-of-principle QRC physical experiments  are ongoing \cite{negoro2018machine,chen2020temporal}. Although all the previous works provide examples of functioning quantum reservoir computers, the fundamental issue raised at the beginning remains open: \textit{what conditions must a physical system fulfill  to be a good quantum reservoir computer?} The aim of this Letter is to establish the relation between operation regime of complex computing systems and performance of QRC. 

Networks of interacting spins enable complex dynamics providing a source  of memory needed for temporal tasks in QRC. The time evolution of these systems and the conditions for thermalization have been recently debated in the context of statistical physics.
Indeed isolated quantum many-body systems can display thermalization in local observables,  as  explained by the Eigenstate Thermalization Hypothesis, which can be seen as the manifestation of ergodicity in quantum mechanics \cite{deutsch1991quantum,srednicki1999approach,d2016quantum}. 
 A remarkable case of a dynamical regime in which this hypothesis is violated is  Many-Body Localization (MBL), where  strong disorder causes the emergence of an extensive number of local integrals of motion that break down the thermalization hypothesis \cite{abanin2019colloquium}. Indeed, such conserved quantities make local observables retain memory of  their initial states. Transitions between localization and thermalization manifest in a critical change of a time-averaged order parameter and are referred to as \textit{dynamical phase transitions} \cite{vzunkovivc2018dynamical,abanin2019colloquium}.
 
The different physical mechanisms underlying the presence or the absence of thermalization  deeply influence the  computational capabilities  of the  different dynamical phases.  
For instance, systems presenting MBL can provide quantum memories at finite temperature \cite{huse2013localization} and avoid overheating in Floquet systems \cite{ponte2015periodically,lazarides2015fate}. In  quantum machine learning, MBL can improve the trainability of parameterized quantum Ising chains \cite{tangpanitanon2020expressibility}. Contrariwise,  localization can be computationally  detrimental in quantum annealing \cite{altshuler2010anderson,laumann2015quantum} or quantum random walk algorithms \cite{keating2007localization,schreiber2011decoherence}. 
Our work establishes that optimal information processing capabilities in QRC not only are favored in the ergodic phase but also that the onset of this regime can be particularly advantageous. We uncover the underlying physical mechanisms favoring machine learning and also provide a new computing perspective on dynamical phases.

\textit{-- Reservoir layer and dynamical phases.} We choose as a reservoir a  spin network described by the transverse-field Ising Hamiltonian plus onsite disorder:
 \begin{equation}\label{Eq:H}
    H=\sum^N_{i>j=1}J_{ij}\sigma_i^x\sigma_j^x+\frac{1}{2}\sum^N_{i=1}(h+D_i)\sigma_i^z,
\end{equation}
where $N$ is the spin number (from now on we will fix $N=10$), $h$ is the  magnetic field, $D_i$ is the onsite disorder, $\sigma^{a}_i$ ($a=x,y,z$) are the  Pauli matrices and $J_{ij}$ are  the spin-spin couplings, randomly selected  from a uniform distribution in the interval $[-J_s/2,J_s/2]$ as it is often done in QRC \cite{fujii2017harnessing,nakajima2019boosting,chen2019learning}. $D_i$ will be also randomly drawn from the uniform distribution $[-W,W]$, where $W$ is the disorder strength. For convenience, all the parameters will be expressed in units of $J_s$.
  
To characterize the dynamical phases of Eq.~(\ref{Eq:H}),  we  use as usual the ratio of adjacent gaps $r_n=\min[\delta_{n+1},\delta_{n}]/\max[\delta_{n+1},\delta_{n}]$, where the gaps are $\delta_n \equiv E_n - E_{n-1}$, and  $\{E_n\}$ is the sorted list in ascending order of the Hamiltonian eigenvalues~\cite{oganesyan2007localization}. We compute the eigenvalues via exact diagonalization limited to one symmetry sector, as the model possesses a parity ($\mathcal{Z}_2$) symmetry \cite{suppmat}.  
 In the localized phase, the level spacing is expected to display a Poisson distribution with $\braket{r}\simeq 0.386$, while in the ergodic phase, according to the random matrix theory, $\braket{r}\simeq 0.535$~\cite{atas2013distribution}.

\begin{figure}[t]
\includegraphics[trim=0cm 0cm 0cm 0cm,clip=true,width=0.45\textwidth]{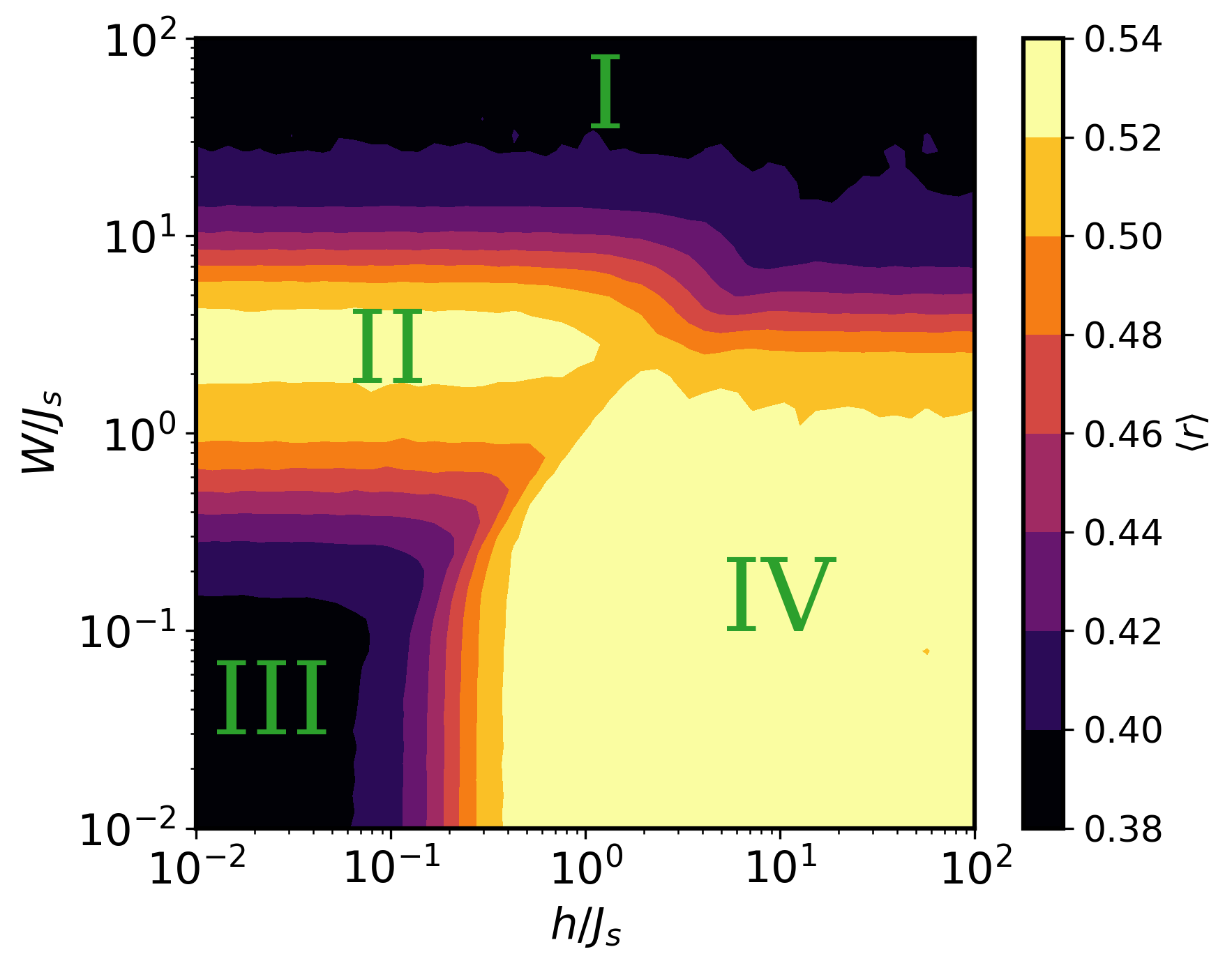}
\caption{ Heat map of $\braket{r}$ for different values of the  magnetic field $h$ and the disorder strength  $W$ in units of $J_s$. Results are averaged over 1200 realizations. 
 }\label{Fig1}
\end{figure}
The full dynamical phase diagram
of  Eq.~(\ref{Eq:H}) depending on magnetic field strength and disorder is shown in Fig.~\ref{Fig1}, displaying four different regions:  two   localization areas  (regions I and III, black) and two  ergodicity areas (regions II and IV, bright yellow). 
The localized regime  in I corresponds to a MBL paramagnetic phase where eigenvalue statistics is Poissonian, while region III corresponds to a  spin-glass phase, where eigenvalue statistics is also Poissonian, but it presents a mobility edge \cite{baldwin2017clustering}.
These regimes have been described in different models as for the transitions III-IV
  \cite{baldwin2017clustering,mukherjee2018many,rademaker2019bridging}, 
 IV-I \cite{maksymov2020many,smith2016many}, or between  different localization phases I-II-III  \cite{huse2013localization,moudgalya2020perturbative,sahay2020emergent}. While such transitions are strictly found in the thermodynamic limit \cite{ khemani2017,scardicchio_epl2020},  signatures are already evident for finite-size systems.

\textit{-- Quantum reservoir dynamics.} 
 The QRC algorithm
 can be divided  into three steps associated to the relative system layers: (i) feed an \textit{input} into the dynamical system; (ii) let the \textit{reservoir},  i.e. the spins network (\ref{Eq:H}),  evolve; (iii)  extract  information from the reservoir, using all or some of its degrees of freedom via an \textit{output layer} \cite{lukovsevivcius2009reservoir}.
Let us assume that our input is given by a sequence $\{s_0,s_1,\dots,s_k,\dots\}$ that is  injected into the same spin (named qubit $1$ for convenience) every time step $k$  \cite{fujii2017harnessing}. 
This spin state is updated every $\Delta t$ as follows: $\rho_1^{(k)}=\ket{\psi_{s_k}}\bra{\psi_{s_k}}$, where $\ket{\psi_{s_k}}=\sqrt{1-s_k}\ket{0}+\sqrt{s_k}\ket{1}$, with $s_k\in [0,1]$. The completely positive, trace-preserving map  summarizing the input encoding and information processing is
\begin{equation}\label{eq:map}
\rho(k\Delta t)=e^{-iH\Delta t}\rho_1^{(k)} \otimes \text{Tr}_1\left\{\rho[(k-1)\Delta t]\right\}e^{iH\Delta t},
\end{equation}
where $e^{-iH\Delta t}$ is the operator of the unitary dynamics and $\text{Tr}_1\left\{\cdot\right\}$ denotes the partial trace performed over the first qubit. The output layer will be built using some of the observables of the system such as the projections $\braket{\sigma^z_i}$ of each spin over the $z$-axis or the spin correlations $\braket{\sigma^z_i\sigma^z_j}$ (Sect. V in \cite{suppmat}). 

\begin{figure}[t]
\includegraphics[trim=0cm 0cm 0cm 0cm,clip=true,width=0.45\textwidth]{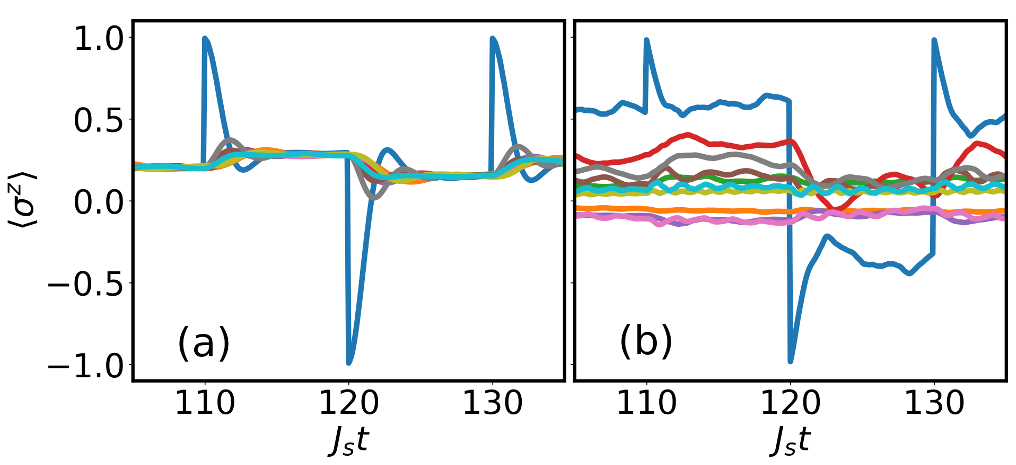}
\caption{ Dynamics of observables $\braket{\sigma^z_i}$ with a binary input ($s_k=\{0,1\}$). Parameters are (a) $W=0$, and $h/J_s=10$; (b) $W/J_s=10$ and $h/J_s=1$. We use $J_s\Delta t=10$ here and in all the next figures. Input is fed to the first spin (blue line) and the rest of lines correspond to the other spins. The initial condition is a random density matrix in both (a) and (b).}
\label{Fig2}
\end{figure}
The spin network response to the input injection through the dynamics of the observables $\braket{\sigma^z_i}$ provides  insightful evidence as shown in Fig.~\ref{Fig2}. The evolution of the observables in Fig.~\ref{Fig2} (a) corresponds to the ergodic region (IV in Fig.~\ref{Fig1}). Color lines represent the different spins, being the input qubit the blue line. This plot displays that all spin observables are driven to an input-dependent stationary state within each $\Delta t$. The explanation behind this behaviour is based on the role of the conserved quantities of the system. In the ergodic regions, total energy $\braket{H}$ and parity $\braket{P}=\braket{\prod^N_i\sigma^z_i}$ are the only conserved quantities and both of them are delocalized. As detailed in Sect. II of \cite{suppmat}, their values after the $k$th input injection ($\braket{H(k\Delta t)}$ and $\braket{P(k\Delta t)}$) only depend on the initial condition of the system $\rho_0$ and the input history up to $s_k$. Then, provided that the unitary dynamics is applied during a time $\Delta t$ allowing for the Eigenstate Thermalization, all $local$ observables, like $\braket{\sigma^z_i}$ in Fig.~\ref{Fig2} (a), only depend on $\braket{H(k\Delta t)}$ and $\braket{P(k\Delta t)}$ up to finite-size fluctuations \cite{abanin2019colloquium}. Therefore, output observables become functions of the input history through $\braket{H}$ and $\braket{P}$, as for the driven dynamics of Fig.~\ref{Fig2} (a). We anticipate that, resetting the first spin state with the input protocol defined above implies a partial information erasure. Repeating this operation several times amounts to losing all trace of the initial conditions as for the convergence property addressed in the following.

Figure~\ref{Fig2} (b) corresponds to the transition from ergodic to MBL (regions IV and I respectively) and displays a significant change of the system response with respect to the previous ergodic case. The observables show now little correlation with respect to the dynamics of the first spin (blue line). This behavior becomes more evident deep in the localized regimes, where none of the $\braket{\sigma^z_i}$ are driven by the input, being instead determined by the initial condition (\cite{suppmat}, e.g. Fig. S2 (d)). The physical reason is that the presence of an extensive number of local conserved quantities hinders the information transport across the network. As a prominent effect of MBL, only those conserved quantities involving the first spin are modified by the input, while all the others keep memory of their initial conditions.

\textit{-- Convergence.}
A fundamental property a system must exhibit to serve as RC is the convergence or echo state property \cite{jaeger2001echo}. This means that, after repeated input injection, the reservoir forgets its initial condition. This is closely related to the so-called fading memory, that is the ability of the output variables to only depend on the recent history of the input sequence \cite{grigoryeva2018echo}.
The convergence property is captured by the distance between two different reservoir states after several (here $200$) input injections through the protocol in Eq.~(\ref{eq:map}). The initial conditions considered are two random density matrices with a typical distance around $||\rho_A-\rho_B||\sim 0.044$ measured using the Frobenius norm, defined as $||A||=\sqrt{\text{Tr}(A^\dagger A)}$ \cite{suppmat}.
\begin{figure}[t]
\includegraphics[trim=0cm 0cm 0cm 0cm,clip=true,width=0.42\textwidth]{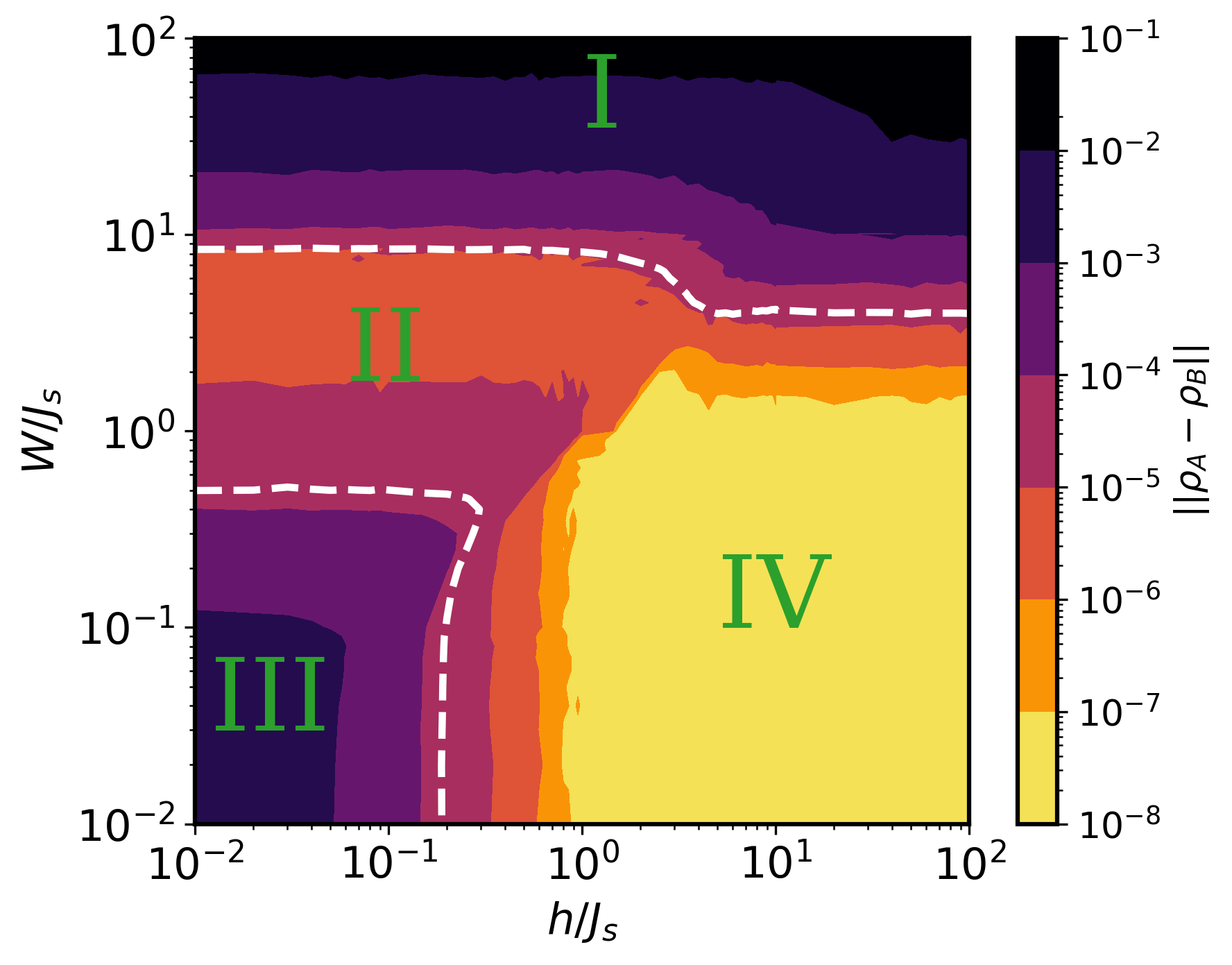}
\caption{Convergence of the system for two different (random) initial conditions after 200 inputs. Distance values are averaged over $600$ realizations, and those below the threshold of $10^{-8}$ are kept to this minimum value for clarity. The white-dashed line corresponds to an intermediate value of $\braket{r}=0.46$ in Fig.~\ref{Fig1}, used to guide the eye. }\label{Fig3}
\end{figure}
Convergence as a function of $h$ and $W$ is shown  in Fig.~\ref{Fig3} and exhibits an insightful correspondence with the phase diagram in Fig.~\ref{Fig1}. Indeed,
the convergence property is enhanced in the ergodic phase enabling then the realization of QRC with the spins reservoir in this regime. The influence of different initial conditions persists in the localized phases, which hinders the QRC performance as it will be shown later. Even if the map~\eqref{eq:map} is contractive due to the partial trace, the local conserved quantities make the contractiveness of this dynamical map much weaker than in the ergodic phase. Indeed, in the MBL phase the creation of entanglement between the first qubit and the rest of the network is very weak, as it only grows logarithmically in time \cite{abanin2019colloquium}. This provides a different focus of dynamical phases, in terms of their response to external perturbations.

\textit{-- RC performance.}\label{sec:performance}
While  convergence  can be seen as a necessary condition to identify suitable RC systems, a characterization of the information processing capabilities of the spin network  is needed in order to determine if the reservoir computer can accomplish a given task. A convenient known advantage of RC is that the output layer is the only one that needs to be trained, by optimizing a linear combination of the responses of the reservoir to the task at hand \cite{lukovsevivcius2009reservoir}.  Given the set of observables $\textbf{x}_{k}$ chosen as output, we write the output layer as ${y_k=\textbf{w}^{\top}\textbf{x}_{k}}$, where $\textbf{w}$ are the weights that are adapted  by minimizing the error with respect to a target function $\bar{y}_k$. Training of the output weights $\textbf{w}$ is usually  done using a linear regression \cite{lukovsevivcius2009reservoir,suppmat}. 

To evaluate the reservoir performance, we will consider two different specific tasks as well as a more general indicator for the processing capacity \cite{dambre2012information}. Let us start with the nonlinear auto-regressive moving average (NARMA) model, which is widely used  to characterize recurrent neural networks \cite{atiya2000new}. The general NARMA$n$ task is defined as:
\begin{equation}\label{eq:NARMA}
   \bar{y}_{k}=0.3\bar{y}_{k-1}+0.05\bar{y}_{k-1}\left(\sum^{n}_{j=1}\bar{y}_{k-j} \right) +1.5 s_{k-n}s_{k-1}+0.1,
\end{equation}
where $n$ is the maximum delay, $\bar{y}_k$ is the target and $s_k$ is a random input uniformly distributed, as detailed in Sect. VI of \cite{suppmat}. For this task, the quantum reservoir computer needs to learn to emulate Eq.~(\ref{eq:NARMA}) from the random input $s_k$, i.e. to reproduce a quadratic nonlinear function of the input sequence up to a maximum fixed delay. We fix the maximum delay of Eq.~\eqref{eq:NARMA} to $n=10$. In order to address performance both in nonlinear and linear temporal tasks, we also evaluate the linear temporal task  $\bar{y}_k=s_{k-\tau}$ fixing $\tau=10$. The value $10$ in both tasks sets the target memory: indeed the reservoir needs a memory of at least 10 past inputs at each time step $k$ to solve them.
The performance of the spin network for each task can be measured as $  C={\text{cov}^2(\bar{\textbf{y}},\textbf{y})}/[{\sigma^2(\textbf{y})\sigma^2(\bar{\textbf{y}})]}$
where $\text{cov}$ is the covariance,  $\sigma^2$ is the variance and $\textbf{y}$ and $\bar{\textbf{y}}$ are the time series of prediction and target respectively. The performance $C$ is bounded between $0$ (target not approximated, the system is useless for this task) and  $1$ (perfect match between prediction and target).

\begin{figure}[htb]
\includegraphics[width=\columnwidth]{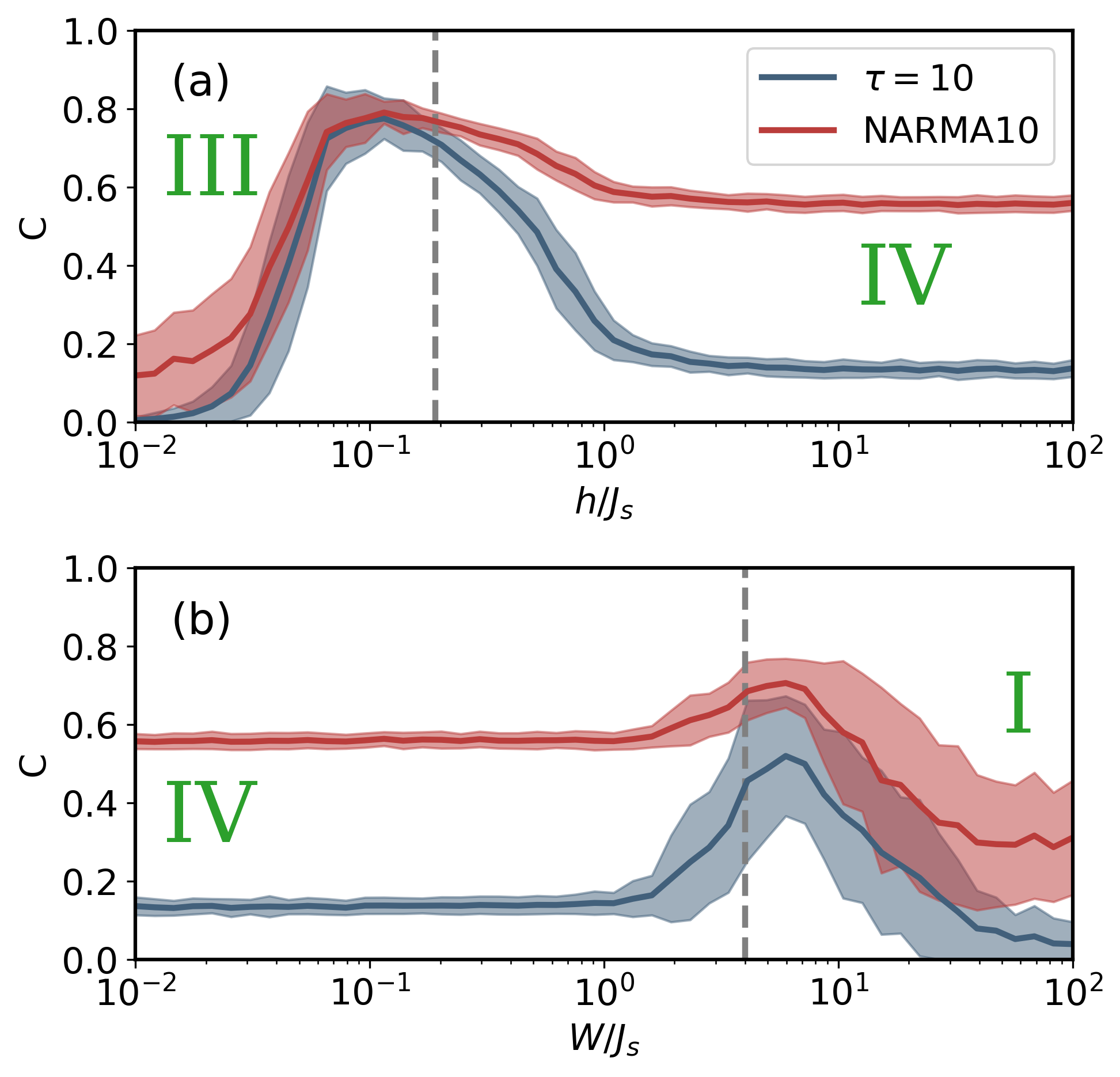}
\caption{Performance covariance $C$ for the NARMA10 and linear memory $\tau=10$ tasks versus $h$ (a) and $W$ (b). We took $W=0$ for (a) and $h/J_s=10$ for (b). We represent the average value of $C$ over 100 realizations of the random network with a solid line, while the shadows represent the standard deviation. The grey-dashed line corresponds to the  intermediate value $\braket{r}=0.46$ in Fig.~\ref{Fig1}, used to guide the eye. The initial condition for all the realizations is the maximal coherent state $\rho_0=\frac{1}{2^N}\sum\ket{i}\bra{j}$.}\label{Fig4}
\end{figure}

In this work, we benefit from the high-dimensionality of the Hilbert phase space of the quantum reservoir by considering an output layer of $O=75$ observables (going beyond classical RC with one-body observables): all the local spin projections $\braket{\sigma^x_j}$, $\braket{\sigma^y_j}$ and $\braket{\sigma^z_j}$ $(1\leq j\leq N)$, plus the two-spin  correlations  along the $z$-axis $\braket{\sigma^z_i\sigma^z_j}$  $(1\leq i,j\leq N$, $i\neq j$). The measurement of these observables is experimentally feasible (Sect. V in \cite{suppmat}). Figure~\ref{Fig4} displays the figure of merit  $C$ for the performance of the spin-based reservoir in the NARMA10 and linear memory tasks. In Fig.~\ref{Fig4} (a), we plot  $C$  versus $h$ keeping $W=0$, while in Fig.~\ref{Fig4} (b) we fix $h/J_s=10$ and vary $W$, in order to detect the change in performance at the transition between  different dynamical phases. Localized regimes (regions I and III) show the smallest values of $C$: the ability to reproduce the target is poor due to their slow convergence properties and actually can be influenced both by the initial elapsed time and the specific choice of initial conditions for the reservoir. In contrast, the ergodic regime (region IV) shows a higher performance achieved rapidly and independently of the reservoir initial state.

Interestingly,  in both plots $C$ shows a peak at the transitions between ergodic and localized phases.
 To show the generality of our results beyond specific tasks and to shed light on the possible performance enhancement at the phase transition, we evaluate \cite{suppmat} the information processing capacity (IPC) \cite{dambre2012information}. We find 
 that the linear memory builds up first as we move away from the localized phase into the ergodic one. Deep into the ergodic region, nonlinear memory dominates. This trade-off between linear and nonlinear memory brings the performance enhancement at the transition for the NARMA$10$ task. Although the IPC results indicate  that one could find nonlinear tasks where the optimal working point is found in the ergodic region and not at the transition, it is often the case that RC tasks precisely require a combination of linear memory and nonlinearity, as often reported at the edge of stability between different dynamical regimes \cite{carroll2020reservoir}.

\textit{ -- Conclusions.}  
High performance in QRC can be achieved thanks to the large dimensionality of the Hilbert space. 
Still, the performance of a system to be used as a quantum reservoir computer crucially depends on its operation regime. We showed in this work that localization, because of the presence of local conserved quantities, is detrimental for an optimal information processing performance due to a slow convergence \cite{lackconv}. We demonstrate this in specific tasks as well as in the quantification of the reservoir memory through the IPC. In contrast, the ergodic phase offers a suitable scenario for the convergence property and facilitates  efficient information extraction. Different tasks can be solved by exploiting the trade-off between linear and nonlinear memory at the phase transition, and actually the onset of thermalization can be particularly advantageous for QRC, a feature reminiscent of the performance enhancement found in classical RC at the edge of stability \cite{carroll2020reservoir}. Our QRC study offers an original perspective on thermal and localized phases in terms of their ability to process information and can be further explored in the context of  quantum correlations, information scrambling, OTOC (out-of-time-order correlators) \cite{swingle2018unscrambling} and transient
real-time evolution of Loschmidt echoes \cite{heyl2013,vzunkovivc2018dynamical}.

Furthermore, our results define the proper conditions for experimental implementations of QRC. We showed the importance of tuning the reservoir at the onset of thermalization, which can be easily achieved by controlling the (average) strength of the magnetic field. Another relevant issue concerns  the network topology. It is already accepted that random connections are necessary for an optimal performance, avoiding redundancies between different degrees of freedom. But it is not enough. We show that even those topologies with disorder leading to an extensive number of conserved quantities are not suited for RC. Strategies for on-line data processing addressing quantum measurement \cite{chen2020temporal,eddins2019high}, need to be further explored \cite{suppmat}. First experiments involve ensemble computing,  obtained by taking many copies of the reservoir \cite{negoro2018machine}, or rely on the use of non-demolition measurements \cite{chen2020temporal}. Several platforms, ranging from trapped-ion quantum simulators \cite{smith2016many,zhang2017observation,PhysRevLett.125.120605}, to optical lattices \cite{Choi1547,Kaufman794} to superconducting circuits \cite{Houck2012,PhysRevLett.120.050507}, or photonic simulators \cite{pierangeli2019large,pierangeli2020scalable} are mature to establish the potential of QRC towards applications, both for classical and quantum time series processing \cite{mujal2021opportunities}.

\begin{acknowledgments}
We acknowledge the Spanish State Research Agency, through the Severo Ochoa and Mar\'ia de Maeztu Program for Centers and Units of Excellence in R\&D (MDM-2017-0711) and through the  QUARESC project (PID2019-109094GB-C21 and -C22/ AEI / 10.13039/501100011033);
 We also acknowledge funding by CAIB through the QUAREC project (PRD2018/47). 
Part of this work has been funded by MICINN/AEI/FEDER and the University of the Balearic Islands through a ``Ramon y Cajal'' Fellowship (RYC-2015-18140) for MCS and a predoctoral  fellowship (MDM-2017-0711-18-1) for RMP. GLG is funded by the Spanish  Ministerio de Educaci\'on y Formaci\'on Profesional/Ministerio de Universidades   and  co-funded by the University of the Balearic Islands through the Beatriz Galindo program  (BG20/00085).
\end{acknowledgments}

\onecolumngrid

\section*{Supplemental Material}
\setcounter{section}{0}
\section{Numerical methods}\label{sec:numerical}
To characterize the dynamical phases of the system's Hamiltonian $ H=\sum^N_{i>j=1}J_{ij}\sigma_i^x\sigma_j^x+\frac{1}{2}\sum^N_{i=1}(h+D_i)\sigma_i^z$ (Fig.~1 of the main text), we employ the ratio of adjacent gaps $r_n=\min[\delta_{n+1},\delta_{n}]/\max[\delta_{n+1},\delta_{n}]$, where the gaps are $\delta_n \equiv E_n - E_{n-1}$, and  $\{E_n\}$ is the sorted list in ascending order of the Hamiltonian eigenvalues~\cite{oganesyan2007localization}. The transition from ergodic (thermalizing) to non-ergodic regions in the parameter space can be tracked down looking at the eigenvalue statistics: in regions where the system has an extensive number of integrals of motion, the eigenvalues  belonging to different sectors can be seen as independent random variables, so that they are spaced according to a Poisson distribution ($\braket{r}\sim 0.386$, where $\langle\rangle$ is the average over all the gaps); on the other hand, in the presence of thermalization, there is level repulsion and the statistics follows the Wigner-Dyson distribution ($\braket{r}\sim 0.53$).
The eigenvalues of the Hamiltonian are computed via exact diagonalization. For a system of 10 spins this implies to diagonalize a matrix of $\sim 10^6$ elements obtaining a sequence of about 1000 eigenvalues. It is convenient that as the model possesses a parity ($\mathcal{Z}_2$) symmetry, the eigenvalue calculation is limited to one symmetry sector. We have employed the QuSpin library for this purpose \cite{weinberg2016quspin}. 

When considering this system as the reservoir for learning purposes,
the dynamics is instead governed by the  dynamical map we repeat here 
\begin{equation}\label{eq:map2}
\rho(k\Delta t)=e^{-iH\Delta t}\rho_1^{(k)} \otimes \text{Tr}_1\left\{\rho[(k-1)\Delta t]\right\}e^{iH\Delta t},
\end{equation}
being $\rho_1$ the density matrix of the first spin (where we introduce the input, see main text),  $e^{-iH\Delta t}$  the operator of the unitary dynamics and $\text{Tr}_1\left\{\cdot\right\}$ denotes the partial trace performed over the first qubit. 
This completely positive and trace preserving (CPTP) map of the QRC algorithm is applied computing again the Hamiltonian eigenvalues via exact diagonalization, taking the whole spectrum this time. In this way, we obtained the results of Figs.~2, 3 and 4 of the main text.
Some elements of the code were generated with the help of the QuTiP library \cite{johansson2012qutip}, such as the random initial density matrices of Fig.~3 of the main text.

\section{Role of ergodicity}
In order to have a theoretical insight about the results presented in the main text, let us analyze how  information enters and leaves the system.
First, let us explain how the input protocol affects the overall picture.
After each input injection, classical information about the input is encoded in the first spin, which instantaneously modifies the expected value of operators with support in this spin. For example, observables like $\braket{\sigma^z_1}$, $\braket{\sigma^x_1\sigma^x_2}$ or the total energy $\braket{H}$ (which has support over all qubits) are instantaneously changed by the input protocol. In particular, those observables with support on the first spin that are conserved quantities become functions of both the initial state of the system and the input sequence, since the unitary dynamics do not affect them. In the case of the total energy $\braket{H}$, this dependence can be expressed as:
\begin{equation} \label{Eq:E}
    \braket{H(k\Delta t)}=\text{Tr}[H \rho(k\Delta t)]=E(\rho_0,s_0\dots,s_{k-1},s_k),
\end{equation}
where $\rho_0$ is the initial state of the system and $s_k$ is the input at time step $k$. 
In the ergodic regions, total energy $\braket{H}$ and parity $\braket{P}=\braket{\prod^N_i\sigma^z_i}$ are the only conserved quantities, which are global operators with support over all the qubits. Provided that the unitary dynamics is applied during a time $\Delta t$ long enough, the Eigenstate Thermalization Hypothesis  guarantees that all \textit{local} observables $\mathcal{O}$ only depend on $\braket{H(k\Delta t)}$ and $\braket{P(k\Delta t)}$ up to finite-size fluctuations \cite{abanin2019colloquium}. 
Indeed,  after $k$ input injections, the expected value of any of such local observables can be well approximated as
\begin{equation}\label{eqr1}
   \langle \mathcal{O}(k \Delta t)\rangle\simeq \frac{{\rm Tr} [\mathcal{O}e^{-\beta_k H-\gamma_k P}]}{\mathcal{Z}_k} ,
\end{equation}
   with the partition function $\mathcal{Z}_k={\rm Tr} [e^{-\beta_k H-\gamma_k P}]$   and where the Lagrange multipliers  $\beta_k$ and $\gamma_k$ can in principle be calculated respectively as
   \begin{eqnarray}
   \langle H(k \Delta t)\rangle=-\frac{\partial \log\mathcal{Z}_k}{\partial \beta_k},\\
    \langle P(k \Delta t)\rangle=-\frac{\partial \log\mathcal{Z}_k}{\partial \gamma_k}.
    \end{eqnarray}

Then, local observables become functions (each of them with its own functional dependence) of the input history through $\braket{H}$ and $\braket{P}$.
\begin{figure}[htb]
\includegraphics[trim=0cm 0cm 0cm 0cm,clip=true,width=0.95\textwidth]{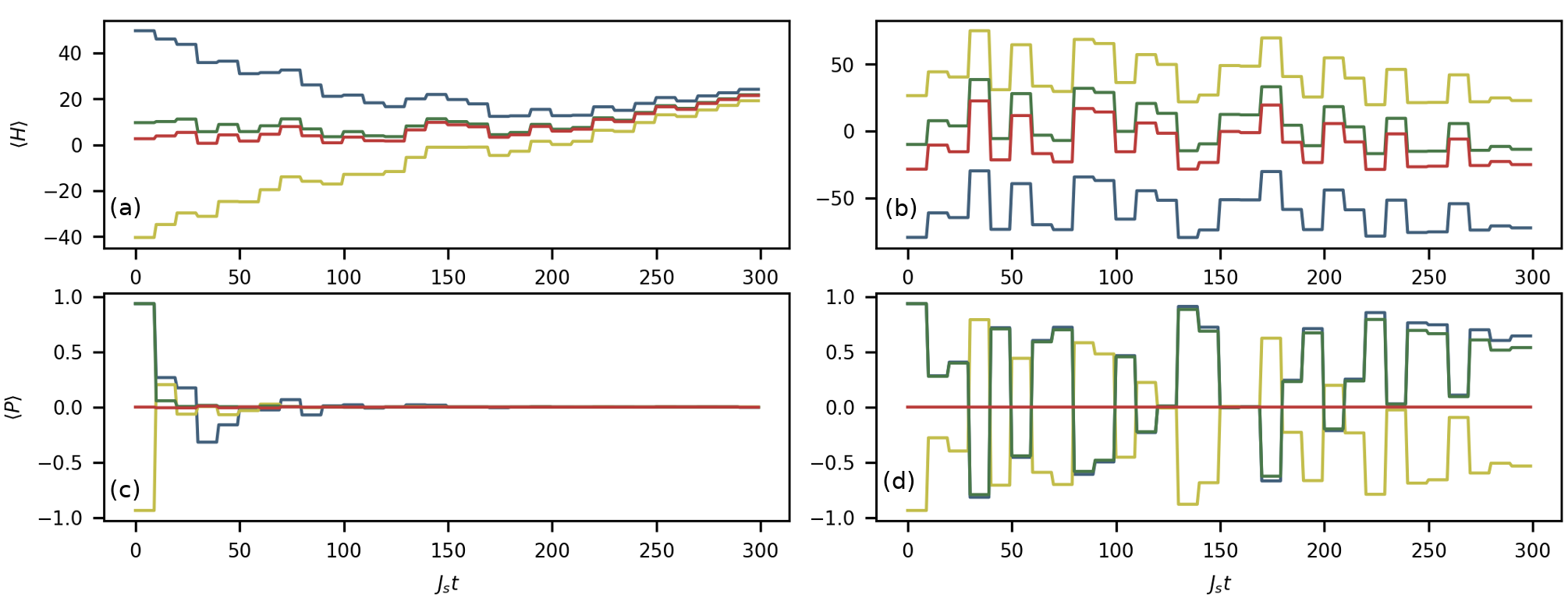}
\caption{Single realizations of the energy $\braket{H}$ and parity $\braket{P}$ dynamics under the same input sequence with different initial conditions. Parameters are $W=0$ and $h/J_s=10$ (ergodic region) for (a) and (c), and $W/J_s=100$ and $h/J_s=1$ (MBL region) for (b) and (d). The number of spins is $N=10$ and the time between inputs is $J_s\Delta t=10$. The color lines correspond to different initial conditions: all spins up in the $z$ basis (blue), all spins down in the $z$ basis (red), half spins up and half spins down in the $z$ basis (green), half spins up and half spins down in the $x$ basis (yellow). These particular initial conditions have been chosen for the sake of visualization.}\label{FigEP}
\end{figure}

In Fig.~\ref{FigEP}, we show an example of the dynamics of $\braket{H(k\Delta t)}$ and $\braket{P(k\Delta t)}$ in the ergodic phase for different initial conditions but the same input sequence. As it can be seen in Figs.~\ref{FigEP} (a) and (c), all the trajectories converge to one single trajectory, which is just a consequence of the convergence property studied in the main text. What we can learn here is that the partial trace operation (see Eq.~\ref{eq:map2}) that accompanies each input injection implies a partial information erasure, as the presence of global conserved quantities implies a significant amount of entanglement between the first spin and the rest of the network. Repeating this operation enough times amounts to losing all trace of the initial conditions and  inputs in the distant past, which ensures the convergence observed in Fig.~3 of the main text. Then, after several input injections, energy and parity can be expressed as
\begin{equation}\label{Eq:E2}
\begin{split}
    &\braket{H(k\Delta t)}\simeq E(s_{k-\alpha},\dots,s_k),\\
    &\braket{P(k\Delta t)}\simeq p(s_{k-\alpha},\dots,s_k),
\end{split}
\end{equation}
where both $E$ and $p$ are only functions of the input history up to a time step $\alpha$ in the past. As a consequence of Eq. (\ref{Eq:E2}), our reservoir is equipped with the so-called fading memory property, which means that two input sequences that are close in the recent past produce outputs that are close in the present.
This fading memory  observed in the collective conserved quantities $\braket{H}$ and $\braket{P}$ is responsible for the good functioning of the reservoir observed in Fig.~4 of the main text. Indeed,  the collectiveness of these degrees of freedom propagates such fading memory to all  degrees of freedom of the reservoir through Eq. (\ref{eqr1}).

The underlying physics is completely different for the localized regimes, where the presence of local integrals of motion, that is, of conserved quantities that have support only on parts of the reservoir, hinders the information transport across the network (Fig. \ref{FigEP}, right panels (b) and (d)). Let us consider the Many-Body Localization (MBL)  paramagnetic phase (region I in Fig. 1 of the main text). As in the ergodic phases, local observables in a MBL phase arrive at a stationary state after a transient time of the unitary dynamics \cite{abanin2019colloquium}. The main difference with respect to the ergodic phase is that now the stationary value of local observables is deeply influenced by the conserved quantities with support on them, and will be weakly affected by the partial trace. Then, information erasure will be much less effective, which makes the reservoir unable to efficiently forget its initial state. Once we introduce the input protocol, only the observables with support in the first spin will be significantly affected after each input injection and after each partial trace application.
This slows down the flow of information in the system: the input enters but hardly arrives at the other spins, and information contained in local conserved quantities that are not connected to the first spin does not easily flow back to it and persists in the network. 

As a consequence, from the point of view of the learning algorithm, not all local observable allow for an efficient extraction of the input information and cannot be exploited, so that the performance is deteriorated: deep in the MBL phase, input information is almost exclusively contained in the first spin, and is erased after each partial trace. This effect can be seen in Fig.~\ref{Fig1S} (d).

In the spin-glass phase (region III) we find a mobility edge, but there are arguments in favor of the presence of local conserved quantities: near-integrable systems ($h\ll1$) can present them \cite{ilievski2015complete,vosk2014dynamical}; furthermore, as conjectured in \cite{geraedts2017emergent}, emergent integrability can be extended to systems with mobility edges, obtaining the conserved quantities from the localized eigenstates.

\section{Dynamical response of the spin network reservoir}
In the main text, we characterized the response of our system to the input by showing the dynamics of the observables $\braket{\sigma^z_i}$. For the sake of succinctness, we showed in Fig.~2 the dynamics of the observables that correspond to the ergodic region (Fig.~2 (a)) and the transition from the ergodic phase to MBL paramagnetic  (Fig.~2 (b)). Here we provide an additional characterization with examples in the remaining dynamical phases.

\begin{figure}[h!]
\includegraphics[width=0.9\columnwidth]{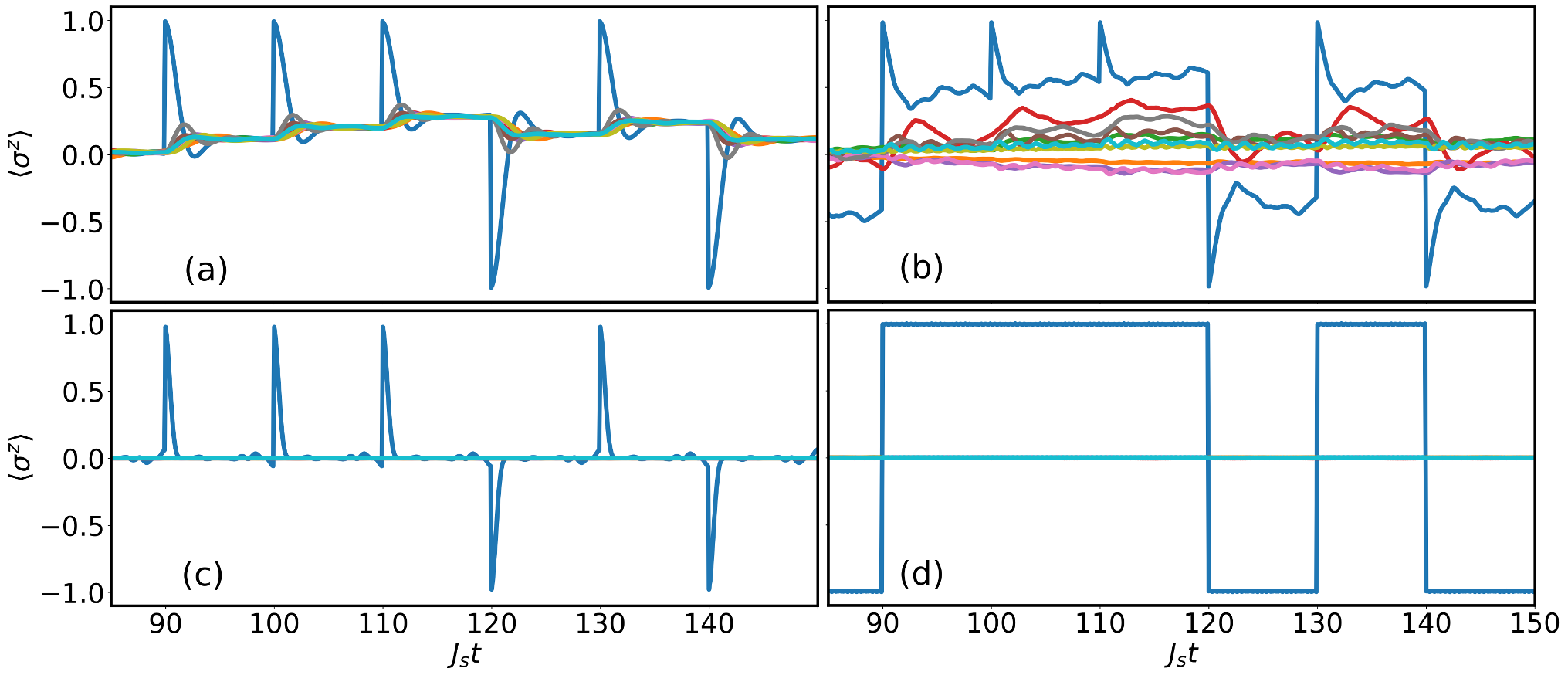}
\caption{Dynamics of observables $\braket{\sigma^z_i}$ with a binary input ($s_k=\{0,1\}$) for $N=10$ spins. Parameters are $J_s\Delta t=10$, (a) $W=0$, and $h/J_s=10$ (region IV); (b) $W/J_s=10$ and $h/J_s=1$ (transition between regions I and II); (c) $W=0$ and $h/J_s=0.01$ (region III); (d) $W/J_s=100$ and $h/J_s=1$ (region I). Input is fed to the first spin (blue line). The initial condition for all the realizations is a random density matrix.}\label{Fig1S}
\end{figure}

Fig.~\ref{Fig1S} presents the dynamics of the observables $\braket{\sigma^z_i}$ as in Fig.~2 of the main text and over a longer time interval. The different lines in Fig.~\ref{Fig1S} correspond to the value of the observable for each spin as a function of time. The blue line represents the dynamics of the first spin, where we inject the input every $\Delta t$. Figs.~\ref{Fig1S} (a) and (b), correspond to the plots as in Fig.~2 of the main text, i.e. we have the parameters $W=0$ and $h/J_s=10$ for (a) and $W/J_s=10$ and $h/J_s=1$ for (b), but with twice the time window in the $x$-axis. Fig.~\ref{Fig1S} (c) represents a point in the spin-glass (region III in Fig.~1 of the main text), with parameters $W=0$ and $h/J_s=0.01$. Fig.~\ref{Fig1S} (d) represents a point in the MBL paramagnetic (region I), with parameters $W/J_s=100$ and $h/J_s=1$.
We note that Fig.~\ref{Fig1S} (a) is a paradigmatic example of a driven system, showing a clear correlation between the dynamics of the observables and the input. However, Figs.~\ref{Fig1S} (b), (c) and (d) present to us cases where the driving of the input has a lesser influence than in Fig.~\ref{Fig1S} (a). 
The fact that some observables in Fig.~2 (b) seem to follow better the trend of the first qubit is a nontrivial point. After analyzing several examples of different reservoirs, with couplings between spins both with random strengths and any/only positive/only negative signs (not shown), our conclusion is that it does not appear to be a local effect but rather a global one, determined by the whole network structure. On the other hand, Figs.~(c) and (d) correspond to the extreme cases where there is no apparent correlation between observables and input. In both cases, we argue that the presence of local integrals of motion is the reason for this loss of correlation between the dynamics of the observables and the input.  In (d), we can identify that localization freezes even the dynamics of the first spin after input injections. On the other hand, Fig.~\ref{Fig1S} (c) also shows a case where the rest of the spins do not respond to the input injection. In contrast, the dynamics of the whole network depends on the initial condition (choosing different initial conditions, not shown, leads to completely different trajectories of the observables). Indeed, local conserved quantities involving the first spin hinder the spreading of information, which explains the lack of response from the other spins.  

\section{On the convergence properties of the different dynamical phases}

In this Section, we tackle the dependence on $\Delta t$ and the number of input injections for the convergence property. The current discussion relates to the results presented in Fig. 3 of the main text. The evolution of Eq.~\eqref{eq:map2} in Sec.~\ref{sec:numerical} is evaluated exactly  for different initial conditions.  First, our numerical results indicate that this completely positive and trace-preserving  map   is strictly contractive (for the considered random initial conditions), i.e. $||\rho_A(k\Delta t)-\rho_B(k\Delta t)||<||\rho_A[(k-1)\Delta t]-\rho_B[(k-1)\Delta t]||$. The only source of contractiveness leading to a convergence of the dynamics after consecutive input injections is the partial trace over the first spin, since the unitary dynamics leaves invariant the distance between density matrices. Then, the more input injections we make (with the associated partial trace), the smaller is the distance between the two initial conditions. Interestingly, as discussed in Sec.~\ref{sec:numerical} and illustrated in Fig.~\ref{Fig2S} below, we find that the dynamical phase plays a determinant role in the convergence. 
\begin{figure}[h!]
\includegraphics[width=0.75\columnwidth]{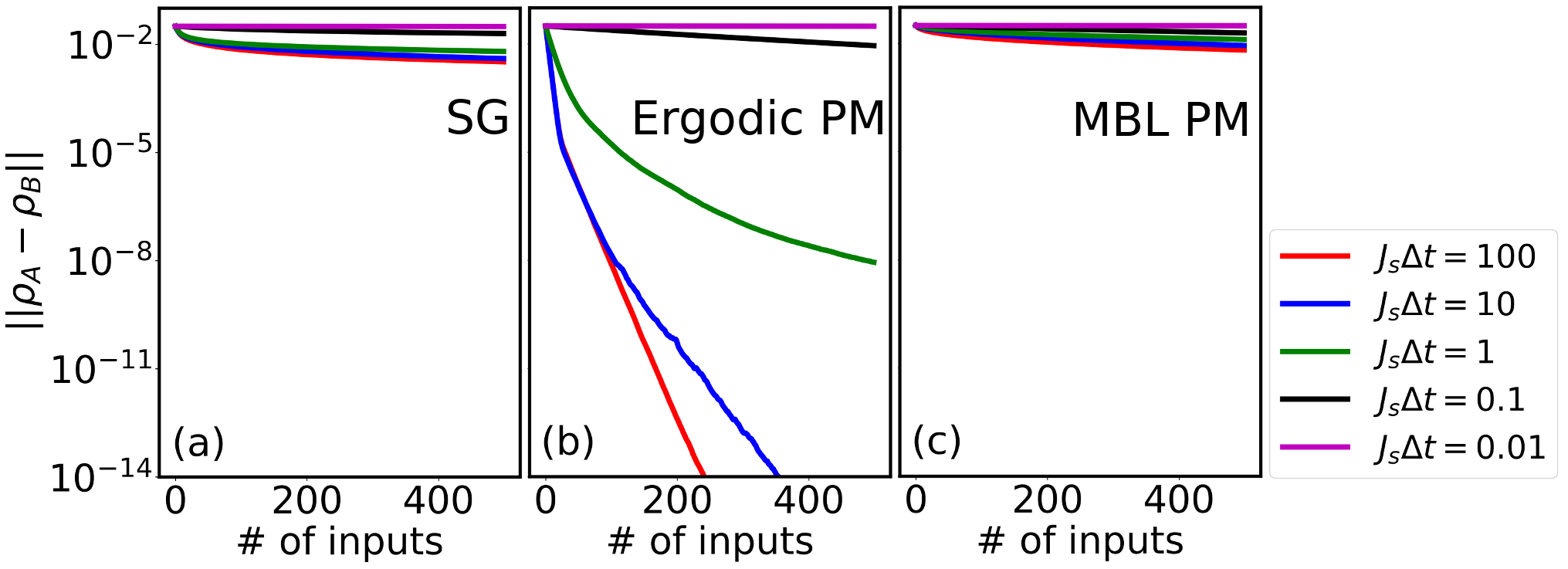}
\caption{Convergence of the system for two different (random) initial conditions over repeated input injection for various periods of injection $\Delta t$. Parameters are (a) $W=0$ and $h/J_s=0.01$ (region III); (b) $W=0$ and $h/J_s=10$ (region IV); (c) $W/J_s=100$ and $h/J_s=1$ (region I). Lines are averaged over 100 realizations. The number of spins is $N=10$. }\label{Fig2S}
\end{figure}

Figs.~\ref{Fig2S} (a), (b) and (c) represent the distance between two random initial conditions over repeated input injection. Figures \ref{Fig2S} (a) and (c) correspond to the spin-glass and MBL paramagnetic phases respectively, while Fig.~\ref{Fig2S} (b) belongs to the ergodic phase. In the three cases, we present the distance for different periods of injection $\Delta t$ such that we are able to show the influence of this parameter in each phase. While for short $\Delta t$ ($J_s\Delta t \leq 0.1$) the results are very similar for all dynamical regimes (the dynamical map is basically the identity), we find a large difference between the behavior of the convergence for the localized regimes and the ergodic phase for large $\Delta t$. Figures \ref{Fig2S} (a) and (c) present curves with a small slope with respect to the number of consecutive injected inputs. In addition, these two cases present a minor dependence on $\Delta t$. In contrast, the value of $\Delta t$ has a strong influence over the convergence property in the ergodic phase, being 200 consecutive input injections more than enough to find a clear convergence when $J_s\Delta t \geq 10$.
This is consistent with the physical behavior characterizing the ergodic phase: after the relaxation time all the spins of the network achieve coherent  coupling. If $\Delta t$ is smaller than this relaxation time,  the convergence is hindered because entanglement does not significantly build-up and information about the initial conditions does not have enough time to flow back to the first spin --and to be erased after each input injection. On the other hand, looking at the localized phases, an extremely slow reduction of the distance between states is reported, making the network in this phase unsuitable for QRC.
We remark that reducing at the minimum the injection time interval is a main requirement in order to process data in RC, being processing speed actually a major performance aspect. We have therefore chosen $J_s\Delta t=10$ as the time interval for our simulations of the main text because it represents a short time scale that still allows relaxation to the thermal state in the ergodic phase, ensuring the validity of Eqs.~\eqref{Eq:E2} above.

\section{Notes on physical implementation}
 
Several comments about the physical implementation of spin-based QRC are in order. First, we describe the construction of the output layer. The $4^N$ observables of a system of $N$ qubits can be defined as $x_j(k\Delta t)=\text{Tr}[B_j \rho(k\Delta t)]=\braket{B_j(k\Delta t)}$, where $B_j$ is the tensor product of $N$ Pauli operators taken from $\{I,\sigma^x,\sigma^y,\sigma^z\}$ (for each qubit). The output layer will be made of some of the observables of the system, such as the projections $\braket{\sigma^z_i}$ of each spin over the $z$-axis or the spins correlations $\braket{\sigma^z_i\sigma^z_j}$. Obtaining these observables is a fundamental question in QRC: we want to harness the large Hilbert space that a quantum system offers, but the processing of information in RC is thought for an online mode. The present approach is based on an ensemble of quantum systems as in \cite{fujii2017harnessing}, where one can disregard the measurement back-action, allowing the online mode computation. Nuclear magnetic resonance experiments, where localized phases  can be observed, \cite{zobov2017effect,wei2018exploring} are suitable for this approach \cite{negoro2018machine}. On the other hand, several platforms, ranging from trapped-ion quantum simulators \cite{smith2016many,zhang2017observation,PhysRevLett.125.120605}, to optical lattices \cite{Choi1547,Kaufman794} to superconducting circuits \cite{Houck2012,PhysRevLett.120.050507}, or photonic simulators \cite{pierangeli2019large,pierangeli2020scalable} are already mature to establish the potential of QRC towards applications. In this case we require the repetition of the experiment to obtain the expected value of observables. But this can be seen as an opportunity to exploit a large number of degrees of freedom. As it is done in ion-trap experimental papers \cite{richerme2014non,zhang2017observation}, expected values of correlations of Pauli matrices can be computed from projective measurements of the individual spins. For example, consider that we want to compute the expected values $\braket{\sigma^z_1}$, $\braket{\sigma^z_3}$ and $\braket{\sigma^z_1\sigma^z_3}$. The expected value of the spin projections are computed with the average of the projective measurements over each spin. For the first spin, one can write this like $\braket{\sigma^z_1}=\frac{1}{M}\sum^M_iz_{1i}$, where $M$ is the number of experimental realizations and $z_{1i}$ accounts for the projection obtained over the first spin for measurement $i$ (a projective measurement produces either $z_{1i}=1$ or $z_{1i}=-1$). One can do the same for the third spin: $\braket{\sigma^z_3}=\frac{1}{M}\sum^M_iz_{3i}$. Finally, to compute the correlation $\braket{\sigma^z_1\sigma^z_3}$ one just needs to use the product of the projections and make the average: $\braket{\sigma^z_1\sigma^z_3}=\frac{1}{M}\sum^M_iz_{1i}z_{3i}$. Thus, it is possible to obtain the correlations in the $z$-axis with the measurements of the spin projections in the $z$-axis, being able to access nontrivial degrees of freedom of our system. We note that one has to repeat more experiments to obtain the projections and correlations in other directions, but there are several papers that propose strategies to surpass this overhead. In particular, there is a recent paper by Huang et al. \cite{huang2020predicting} where they theoretically and numerically demonstrate that one can efficiently estimate a large number of observables (correlations with 2,3,... spins) from a reasonable number of projective measurements, based on the formalism of shadow tomography.

Another fundamental point in QRC is the characteristics of the quantum model simulated by the physical platform. Different experimental platforms will provide different network structures, and then, different couplings in the Hamiltonian. However, the phenomenon of dynamical phase transition in the transverse-field Ising model with long-range interactions of the couplings seems rather universal, either from an order or glassy ground state towards thermalization \cite{zhang2017observation,baldwin2017clustering,mukherjee2018many}, or from the ergodic phase to the MBL phase \cite{smith2016many,maksymov2020many}. We have checked that our results are consistent with the convergence property of other long-range configurations. However, we have not explored the performance of systems without any kind of disorder. We foresee that reducing the number of degrees of freedom with extra symmetries would decrease the performance of the system. On the other hand, topologies with short-range interactions have not been explored either.

\section{Reservoir Computing Training Methodology}

In RC, training of the output weights $\textbf{w}$ is usually done using a linear regression for vector $\textbf{y}=X\textbf{w}$ with respect to $\bar{\textbf{y}}$, where $X$ is a $L\times(O+1)$ matrix that collects the reservoir states at different times, $L$ is the length of the input sequence and $O$ is the  number of output variables, that we remind correspond to expectation values of observables. A constant bias term $x_{k,O+1}=1$ is added for an optimal training \cite{lukovsevivcius2009reservoir}. Sequences of $5000$ inputs are used for Fig.~4 of the main text: $1000$ to  wash out the initial conditions, $2000$ for training, and $2000$ for  prediction, that is to assess the performance in the task. The linear regression is chosen as the training method to offload and simplify the work of the output layer, harnessing the nonlinear processing of information from the reservoir.

\section{NARMA task}
The nonlinear auto-regressive moving average (NARMA) task belongs to a large family of problems that researchers in  the Machine Learning community (especially in the Recurrent Neural Networks community) try to solve. These kinds of tasks consist in approximating a nonlinear and temporal function of the input sequence. The NARMA task in particular consists in approximating a quadratic nonlinear function of the input sequence up to a maximum fixed delay. The general NARMA$n$ model is defined as:
\begin{equation}\label{eq:NARMA2}
   \bar{y}_{k}=0.3\bar{y}_{k-1}+0.05\bar{y}_{k-1}\left(\sum^{n}_{j=1}\bar{y}_{k-j} \right) +1.5 s_{k-n}s_{k-1}+0.1,
\end{equation}
where $n$ is the maximum delay, $\bar{y}_k$ is the target and $s_k$ is a random input uniformly distributed. In the main text, we fix the maximum delay of Eq.~\eqref{eq:NARMA2} to $n=10$, which requires a significant memory capacity, and restrict the input to the interval $[0,0.2]$ to prevent divergences \cite{kubota2019dynamical}. 

\begin{figure}[htb]
\includegraphics[trim=0cm 0cm 0cm 0cm,clip=true,width=0.5\textwidth]{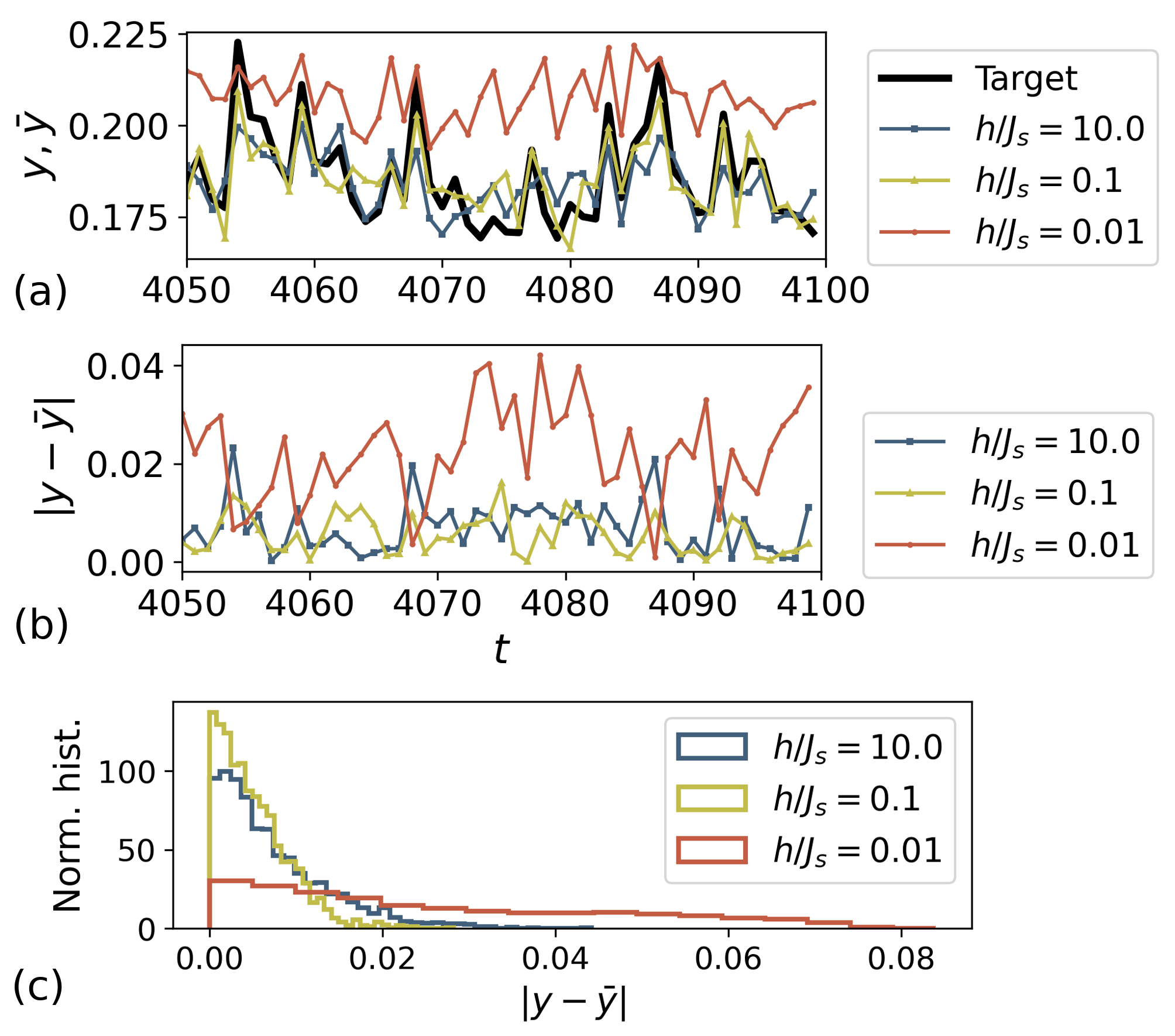}
\caption{Single realizations of the NARMA10 task for different values of $h$. Figure (a) represents a temporal window of the target and output trajectories.  Figure (b) corresponds to the error between target and output reservoirs for the same temporal window, and Fig. (c) is a histogram of all the errors during the test. The total number of time steps was 5000: 1000 for washing out the initial conditions, 2000 for the training and 2000 for the test. Parameters are $N=10$, $W=0$ and $J_s\Delta t=10$. The initial condition for all the realizations is the maximal coherent state $\rho_0=\frac{1}{2^N}\sum\ket{i}\bra{j}$.}\label{FigNARMA}
\end{figure}

To give some intuition about what it means to realize the NARMA task, we show in Fig.~\ref{FigNARMA} some of the results that we obtain for a single realization of the learning algorithm. Figure~\ref{FigNARMA} (a) contains a temporal window of the target and output sequences. The target sequence (black line) represents the trajectory of Eq.~\ref{eq:NARMA2}. The color lines represent the trajectories that the quantum reservoir computer produces as an output after the training process. The selection of parameters corresponds to the three different regions that Fig.~4 (a) of the main text contain: ergodic phase ($h=10$), transition ($h=0.1$) and spin-glass phase ($h=0.01$). As we observe in this figure, the realization of the transition region is the best one approximating the target function, while the realization of the spin-glass phase does not follow the target. Trying to get a better insight, we represent in Fig.~\ref{FigNARMA} (b) the error between the target trajectory and the outputs of the reservoir in the same temporal window. For this time interval, it is clear that $h=0.01$ is the worst case scenario, but  it is less straightforward to assess if $h=0.1$ is better than $h=10$. Then, we need to account for the differences between target and reservoir output trajectories during a larger time interval. In Fig. \ref{FigNARMA} (c) we show a histogram of all the differences obtained between target and outputs during the test sequence (2000 time steps). We observed that the three examples have the maximum of the distribution close to the origin. However, the transition example ($h=0.1$) has the highest maximum with the shortest tail, representing the case with the smallest errors.

\section{On the information processing capacity}

It is impractical to cover the performance of the quantum spin network for all conceivable benchmark tasks. In this spirit, it was recently shown that one can provide a task-independent characterization of the computational capabilities of a dynamical system acting as a reservoir computer by measuring its Information Processing Capacity (IPC) \cite{dambre2012information}. The authors demonstrated that the total computational capacity of a dynamical system is bounded by the number of linearly independent variables that we use for the output. In addition, the computed IPC can only saturate the bound when the dynamical system possesses fading memory, which is strongly connected to the convergence property (or echo state property) and under some conditions (compactness of the input and reservoir state spaces, contractivity and continuity of the reservoir dynamics \cite{grigoryeva2018echo}) both of them are fulfilled simultaneously.

In this context, the computational capabilities of a reservoir are identified as the ability of performing both temporal and nonlinear tasks. Then, computing all of these capabilities would imply evaluating all the possible linear and nonlinear functions of the input sequence that our system can approximate. The original theory developed in \cite{dambre2012information} simplifies this problem by choosing a set of orthogonal functions. The target functions are defined as the product of Legendre polynomials of a given degree $d_i$, such that the sum of the degrees of all the multiplied polynomials add up to a given degree $d\geq 1$ of nonlinearity. Then, the total capacity can be divided into linear ($d=1$) and nonlinear contributions ($d>1$). Besides, the polynomials will be functions of the input at different times in the past. In this way, our targets not only account for nonlinear functions but also temporal maps.  

The capacity coefficient to quantify how well a system reproduces a target function can be defined as in \cite{dambre2012information}:
\begin{equation}
    C_L(X,\textbf{y})=1-\frac{\text{min}_{\textbf{w}}\text{MSE}_L(\textbf{y},\bar{\textbf{y}})}{\braket{\bar{\textbf{y}}^2}_L},
\end{equation}
where $X$ is the matrix of the reservoir variables at different times, $\textbf{y}$ and $\bar{\textbf{y}}$ are the prediction and target sequences respectively and $\textbf{w}$ is the vector of weights of the output layer. The mean square error MSE is the cost function $\text{MSE}_L(\textbf{y},\bar{\textbf{y}})=\frac{1}{L}\sum^L_{k=1}(y_k-\bar{y}_k)^2$ and the bracket $\braket{}_L$ denotes the temporal average for sequences of length $L$.

The target function for a given degree $d$ is:
\begin{equation}\label{Eq:Pol}
    \bar{y}_k=\prod_i \mathcal{P}_{d_i}[\tilde{s}_{k-i}], \quad \sum_{i}d_i=d,
\end{equation}
where $\tilde{s}_{k}$ accounts for the input and $\mathcal{P}_{d_i}$ are the Legendre polynomials of a given degree $d_i$. Due to the choice of polynomials and to preserve their orthogonality, the input must be uniformly distributed over the interval $[-1,1]$, rescaling the inputs introduced in the spin system as $s_k=(1+\tilde{s}_k)/2\in [0,1]$. The numerical simulations have been done with input streams of length $L=10^5$ and maximum degree of nonlinearity $d_{\text{max}}=6$, being sufficient to guarantee the saturation of the total capacity in our case. An initial transient of $10^4$ inputs was taken into account to wash out the initial conditions of the system. Since individual capacities may be slightly overestimated in the numerical analysis, a threshold is set to truncate the smaller contributions. These thresholds have been numerically obtained and depend on the input sequence length $L$, the number of observables in the output layer $O$ and the degree of nonlinearity of the task (see \cite{dambre2012information} for more details).
\begin{figure}[htb]
\includegraphics[trim=0cm 0cm 0cm 0cm,clip=true]{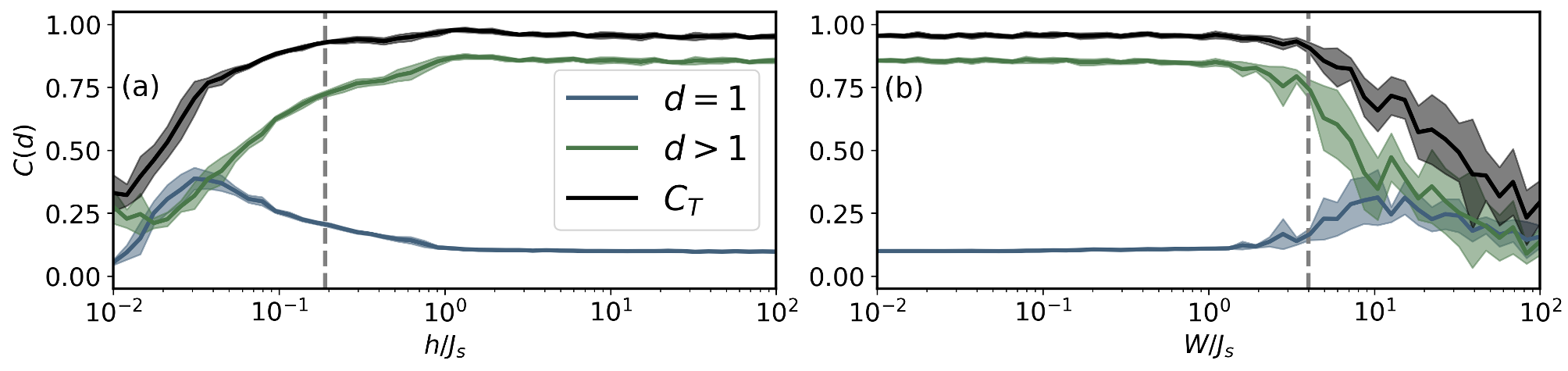}
\caption{Information processing capacity versus $h$ (a) and $W$ (b).  We represent the average value of the total capacity $C_T$ as well as the different contributions by the maximum degree of nonlinearity of the polynomials. Solid lines represent the average values while the shadows are the standard deviation over 10 realizations. Both figures are computed for $N=10$ and $J_s\Delta t=10$, using an output layer of $O=75$ observables (see main text for details). We took $W=0$ for (a) (transition between regions III and IV) and $h/J_s=10$ for (b) (transition between regions IV and I). The grey-dashed line corresponds to the intermediate value $\braket{r}=0.46$ in Fig.~1 in the main text, used to guide the eye. The initial condition for all the realizations is the maximal coherent state $\rho_0=\frac{1}{2^N}\sum\ket{i}\bra{j}$.\label{Fig3S}}
\end{figure}

The results for the normalized IPC are shown in Fig. \ref{Fig3S}. We represent the total capacity of the system, denoted by $C_T$, as well as the linear ($d=1$) and nonlinear ($d>1$) memory contributions to the IPC. We find that the localized regimes (regions I and III) show the smallest values of $C_T$, while the ergodic regime (region IV) shows a stable total capacity close to the maximum value. We also find that linear and nonlinear memory present a trade-off at the transitions between the dynamical phases. The vertical dashed lines in Fig. \ref{Fig3S}, for an intermediate value of $\braket{r}$, interestingly mark the building up of the total memory capacity for increasing magnetic field strength ($h$) and decreasing onsite disorder strength ($W$), respectively. The low IPC values at the localized regimes can be explained with the convergence property: the convergence is so slow that we can assume the loss of the fading memory property for finite input sequences, which is a necessary condition to saturate the total capacity. This slow convergence rate is a consequence of the weak contractiveness of the dynamical map in Eq.~\eqref{eq:map2}: in the MBL phase the creation of entanglement between the first qubit and the rest of the network is very weak (it only grows logarithmically in time), so that the system cannot efficiently dissipate information about inputs injected in the distant past.
\bibliography{references}

\begin{thebibliography}{73}%
\makeatletter
\providecommand \@ifxundefined [1]{%
 \@ifx{#1\undefined}
}%
\providecommand \@ifnum [1]{%
 \ifnum #1\expandafter \@firstoftwo
 \else \expandafter \@secondoftwo
 \fi
}%
\providecommand \@ifx [1]{%
 \ifx #1\expandafter \@firstoftwo
 \else \expandafter \@secondoftwo
 \fi
}%
\providecommand \natexlab [1]{#1}%
\providecommand \enquote  [1]{``#1''}%
\providecommand \bibnamefont  [1]{#1}%
\providecommand \bibfnamefont [1]{#1}%
\providecommand \citenamefont [1]{#1}%
\providecommand \href@noop [0]{\@secondoftwo}%
\providecommand \href [0]{\begingroup \@sanitize@url \@href}%
\providecommand \@href[1]{\@@startlink{#1}\@@href}%
\providecommand \@@href[1]{\endgroup#1\@@endlink}%
\providecommand \@sanitize@url [0]{\catcode `\\12\catcode `\$12\catcode
  `\&12\catcode `\#12\catcode `\^12\catcode `\_12\catcode `\%12\relax}%
\providecommand \@@startlink[1]{}%
\providecommand \@@endlink[0]{}%
\providecommand \url  [0]{\begingroup\@sanitize@url \@url }%
\providecommand \@url [1]{\endgroup\@href {#1}{\urlprefix }}%
\providecommand \urlprefix  [0]{URL }%
\providecommand \Eprint [0]{\href }%
\providecommand \doibase [0]{https://doi.org/}%
\providecommand \selectlanguage [0]{\@gobble}%
\providecommand \bibinfo  [0]{\@secondoftwo}%
\providecommand \bibfield  [0]{\@secondoftwo}%
\providecommand \translation [1]{[#1]}%
\providecommand \BibitemOpen [0]{}%
\providecommand \bibitemStop [0]{}%
\providecommand \bibitemNoStop [0]{.\EOS\space}%
\providecommand \EOS [0]{\spacefactor3000\relax}%
\providecommand \BibitemShut  [1]{\csname bibitem#1\endcsname}%
\let\auto@bib@innerbib\@empty
%</preamble>
\bibitem [{\citenamefont {Jaeger}(2021)}]{jaeger2020exploring}%
  \BibitemOpen
  \bibfield  {author} {\bibinfo {author} {\bibfnamefont {H.}~\bibnamefont
  {Jaeger}},\ }\bibfield  {title} {\bibinfo {title} {Toward a generalized
  theory comprising digital, neuromorphic, and unconventional computing},\
  }\href {http://iopscience.iop.org/article/10.1088/2634-4386/abf151}
  {\bibfield  {journal} {\bibinfo  {journal} {Neuromorphic Computing and
  Engineering}\ } (\bibinfo {year} {2021})}\BibitemShut {NoStop}%
\bibitem [{\citenamefont {Jaeger}(2001)}]{jaeger2001echo}%
  \BibitemOpen
  \bibfield  {author} {\bibinfo {author} {\bibfnamefont {H.}~\bibnamefont
  {Jaeger}},\ }\bibfield  {title} {\bibinfo {title} {The “echo state”
  approach to analysing and training recurrent neural networks-with an erratum
  note},\ }\href@noop {} {\bibfield  {journal} {\bibinfo  {journal} {Bonn,
  Germany: German National Research Center for Information Technology GMD
  Technical Report}\ }\textbf {\bibinfo {volume} {148}},\ \bibinfo {pages} {13}
  (\bibinfo {year} {2001})}\BibitemShut {NoStop}%
\bibitem [{\citenamefont {Maass}\ \emph {et~al.}(2002)\citenamefont {Maass},
  \citenamefont {Natschl{\"a}ger},\ and\ \citenamefont
  {Markram}}]{maass2002real}%
  \BibitemOpen
  \bibfield  {author} {\bibinfo {author} {\bibfnamefont {W.}~\bibnamefont
  {Maass}}, \bibinfo {author} {\bibfnamefont {T.}~\bibnamefont
  {Natschl{\"a}ger}},\ and\ \bibinfo {author} {\bibfnamefont {H.}~\bibnamefont
  {Markram}},\ }\bibfield  {title} {\bibinfo {title} {Real-time computing
  without stable states: A new framework for neural computation based on
  perturbations},\ }\href@noop {} {\bibfield  {journal} {\bibinfo  {journal}
  {Neural Computation}\ }\textbf {\bibinfo {volume} {14}},\ \bibinfo {pages}
  {2531} (\bibinfo {year} {2002})}\BibitemShut {NoStop}%
\bibitem [{\citenamefont {Luko{\v{s}}evi{\v{c}}ius}\ and\ \citenamefont
  {Jaeger}(2009)}]{lukovsevivcius2009reservoir}%
  \BibitemOpen
  \bibfield  {author} {\bibinfo {author} {\bibfnamefont {M.}~\bibnamefont
  {Luko{\v{s}}evi{\v{c}}ius}}\ and\ \bibinfo {author} {\bibfnamefont
  {H.}~\bibnamefont {Jaeger}},\ }\bibfield  {title} {\bibinfo {title}
  {Reservoir computing approaches to recurrent neural network training},\
  }\href@noop {} {\bibfield  {journal} {\bibinfo  {journal} {Computer Science
  Review}\ }\textbf {\bibinfo {volume} {3}},\ \bibinfo {pages} {127} (\bibinfo
  {year} {2009})}\BibitemShut {NoStop}%
\bibitem [{\citenamefont {Appeltant}\ \emph {et~al.}(2011)\citenamefont
  {Appeltant}, \citenamefont {Soriano}, \citenamefont {Van~der Sande},
  \citenamefont {Danckaert}, \citenamefont {Massar}, \citenamefont {Dambre},
  \citenamefont {Schrauwen}, \citenamefont {Mirasso},\ and\ \citenamefont
  {Fischer}}]{appeltant2011information}%
  \BibitemOpen
  \bibfield  {author} {\bibinfo {author} {\bibfnamefont {L.}~\bibnamefont
  {Appeltant}}, \bibinfo {author} {\bibfnamefont {M.~C.}\ \bibnamefont
  {Soriano}}, \bibinfo {author} {\bibfnamefont {G.}~\bibnamefont {Van~der
  Sande}}, \bibinfo {author} {\bibfnamefont {J.}~\bibnamefont {Danckaert}},
  \bibinfo {author} {\bibfnamefont {S.}~\bibnamefont {Massar}}, \bibinfo
  {author} {\bibfnamefont {J.}~\bibnamefont {Dambre}}, \bibinfo {author}
  {\bibfnamefont {B.}~\bibnamefont {Schrauwen}}, \bibinfo {author}
  {\bibfnamefont {C.~R.}\ \bibnamefont {Mirasso}},\ and\ \bibinfo {author}
  {\bibfnamefont {I.}~\bibnamefont {Fischer}},\ }\bibfield  {title} {\bibinfo
  {title} {Information processing using a single dynamical node as complex
  system},\ }\href@noop {} {\bibfield  {journal} {\bibinfo  {journal} {Nature
  Communications}\ }\textbf {\bibinfo {volume} {2}},\ \bibinfo {pages} {1}
  (\bibinfo {year} {2011})}\BibitemShut {NoStop}%
\bibitem [{\citenamefont {Van~der Sande}\ \emph {et~al.}(2017)\citenamefont
  {Van~der Sande}, \citenamefont {Brunner},\ and\ \citenamefont
  {Soriano}}]{van2017advances}%
  \BibitemOpen
  \bibfield  {author} {\bibinfo {author} {\bibfnamefont {G.}~\bibnamefont
  {Van~der Sande}}, \bibinfo {author} {\bibfnamefont {D.}~\bibnamefont
  {Brunner}},\ and\ \bibinfo {author} {\bibfnamefont {M.~C.}\ \bibnamefont
  {Soriano}},\ }\bibfield  {title} {\bibinfo {title} {Advances in photonic
  reservoir computing},\ }\href@noop {} {\bibfield  {journal} {\bibinfo
  {journal} {Nanophotonics}\ }\textbf {\bibinfo {volume} {6}},\ \bibinfo
  {pages} {561} (\bibinfo {year} {2017})}\BibitemShut {NoStop}%
\bibitem [{\citenamefont {Torrejon}\ \emph {et~al.}(2017)\citenamefont
  {Torrejon}, \citenamefont {Riou}, \citenamefont {Araujo}, \citenamefont
  {Tsunegi}, \citenamefont {Khalsa}, \citenamefont {Querlioz}, \citenamefont
  {Bortolotti}, \citenamefont {Cros}, \citenamefont {Yakushiji}, \citenamefont
  {Fukushima} \emph {et~al.}}]{torrejon2017neuromorphic}%
  \BibitemOpen
  \bibfield  {author} {\bibinfo {author} {\bibfnamefont {J.}~\bibnamefont
  {Torrejon}}, \bibinfo {author} {\bibfnamefont {M.}~\bibnamefont {Riou}},
  \bibinfo {author} {\bibfnamefont {F.~A.}\ \bibnamefont {Araujo}}, \bibinfo
  {author} {\bibfnamefont {S.}~\bibnamefont {Tsunegi}}, \bibinfo {author}
  {\bibfnamefont {G.}~\bibnamefont {Khalsa}}, \bibinfo {author} {\bibfnamefont
  {D.}~\bibnamefont {Querlioz}}, \bibinfo {author} {\bibfnamefont
  {P.}~\bibnamefont {Bortolotti}}, \bibinfo {author} {\bibfnamefont
  {V.}~\bibnamefont {Cros}}, \bibinfo {author} {\bibfnamefont {K.}~\bibnamefont
  {Yakushiji}}, \bibinfo {author} {\bibfnamefont {A.}~\bibnamefont
  {Fukushima}}, \emph {et~al.},\ }\bibfield  {title} {\bibinfo {title}
  {Neuromorphic computing with nanoscale spintronic oscillators},\ }\href@noop
  {} {\bibfield  {journal} {\bibinfo  {journal} {Nature}\ }\textbf {\bibinfo
  {volume} {547}},\ \bibinfo {pages} {428} (\bibinfo {year}
  {2017})}\BibitemShut {NoStop}%
\bibitem [{\citenamefont {Tanaka}\ \emph {et~al.}(2019)\citenamefont {Tanaka},
  \citenamefont {Yamane}, \citenamefont {H{\'e}roux}, \citenamefont {Nakane},
  \citenamefont {Kanazawa}, \citenamefont {Takeda}, \citenamefont {Numata},
  \citenamefont {Nakano},\ and\ \citenamefont {Hirose}}]{tanaka2019recent}%
  \BibitemOpen
  \bibfield  {author} {\bibinfo {author} {\bibfnamefont {G.}~\bibnamefont
  {Tanaka}}, \bibinfo {author} {\bibfnamefont {T.}~\bibnamefont {Yamane}},
  \bibinfo {author} {\bibfnamefont {J.~B.}\ \bibnamefont {H{\'e}roux}},
  \bibinfo {author} {\bibfnamefont {R.}~\bibnamefont {Nakane}}, \bibinfo
  {author} {\bibfnamefont {N.}~\bibnamefont {Kanazawa}}, \bibinfo {author}
  {\bibfnamefont {S.}~\bibnamefont {Takeda}}, \bibinfo {author} {\bibfnamefont
  {H.}~\bibnamefont {Numata}}, \bibinfo {author} {\bibfnamefont
  {D.}~\bibnamefont {Nakano}},\ and\ \bibinfo {author} {\bibfnamefont
  {A.}~\bibnamefont {Hirose}},\ }\bibfield  {title} {\bibinfo {title} {Recent
  advances in physical reservoir computing: A review},\ }\href@noop {}
  {\bibfield  {journal} {\bibinfo  {journal} {Neural Networks}\ } (\bibinfo
  {year} {2019})}\BibitemShut {NoStop}%
\bibitem [{\citenamefont {Arute}\ \emph {et~al.}(2019)\citenamefont {Arute},
  \citenamefont {Arya}, \citenamefont {Babbush} \emph {et~al.}}]{Arute2019}%
  \BibitemOpen
  \bibfield  {author} {\bibinfo {author} {\bibfnamefont {F.}~\bibnamefont
  {Arute}}, \bibinfo {author} {\bibfnamefont {K.}~\bibnamefont {Arya}},
  \bibinfo {author} {\bibfnamefont {R.}~\bibnamefont {Babbush}}, \emph
  {et~al.},\ }\bibfield  {title} {\bibinfo {title} {Quantum supremacy using a
  programmable superconducting processor},\ }\href
  {https://doi.org/10.1038/s41586-019-1666-5} {\bibfield  {journal} {\bibinfo
  {journal} {Nature}\ }\textbf {\bibinfo {volume} {574}},\ \bibinfo {pages}
  {505} (\bibinfo {year} {2019})}\BibitemShut {NoStop}%
\bibitem [{\citenamefont {Fujii}\ and\ \citenamefont
  {Nakajima}(2017)}]{fujii2017harnessing}%
  \BibitemOpen
  \bibfield  {author} {\bibinfo {author} {\bibfnamefont {K.}~\bibnamefont
  {Fujii}}\ and\ \bibinfo {author} {\bibfnamefont {K.}~\bibnamefont
  {Nakajima}},\ }\bibfield  {title} {\bibinfo {title} {Harnessing
  disordered-ensemble quantum dynamics for machine learning},\ }\href@noop {}
  {\bibfield  {journal} {\bibinfo  {journal} {Physical Review Applied}\
  }\textbf {\bibinfo {volume} {8}},\ \bibinfo {pages} {024030} (\bibinfo {year}
  {2017})}\BibitemShut {NoStop}%
\bibitem [{\citenamefont {Mujal}\ \emph {et~al.}(2021)\citenamefont {Mujal},
  \citenamefont {Mart{\'\i}nez-Pe{\~n}a}, \citenamefont {Nokkala},
  \citenamefont {Garc{\'\i}a-Beni}, \citenamefont {Giorgi}, \citenamefont
  {Soriano},\ and\ \citenamefont {Zambrini}}]{mujal2021opportunities}%
  \BibitemOpen
  \bibfield  {author} {\bibinfo {author} {\bibfnamefont {P.}~\bibnamefont
  {Mujal}}, \bibinfo {author} {\bibfnamefont {R.}~\bibnamefont
  {Mart{\'\i}nez-Pe{\~n}a}}, \bibinfo {author} {\bibfnamefont {J.}~\bibnamefont
  {Nokkala}}, \bibinfo {author} {\bibfnamefont {J.}~\bibnamefont
  {Garc{\'\i}a-Beni}}, \bibinfo {author} {\bibfnamefont {G.~L.}\ \bibnamefont
  {Giorgi}}, \bibinfo {author} {\bibfnamefont {M.~C.}\ \bibnamefont
  {Soriano}},\ and\ \bibinfo {author} {\bibfnamefont {R.}~\bibnamefont
  {Zambrini}},\ }\bibfield  {title} {\bibinfo {title} {Opportunities in quantum
  reservoir computing and extreme learning machines},\ }\href@noop {}
  {\bibfield  {journal} {\bibinfo  {journal} {arXiv preprint arXiv:2102.11831}\
  } (\bibinfo {year} {2021})}\BibitemShut {NoStop}%
\bibitem [{\citenamefont {Nakajima}\ \emph {et~al.}(2019)\citenamefont
  {Nakajima}, \citenamefont {Fujii}, \citenamefont {Negoro}, \citenamefont
  {Mitarai},\ and\ \citenamefont {Kitagawa}}]{nakajima2019boosting}%
  \BibitemOpen
  \bibfield  {author} {\bibinfo {author} {\bibfnamefont {K.}~\bibnamefont
  {Nakajima}}, \bibinfo {author} {\bibfnamefont {K.}~\bibnamefont {Fujii}},
  \bibinfo {author} {\bibfnamefont {M.}~\bibnamefont {Negoro}}, \bibinfo
  {author} {\bibfnamefont {K.}~\bibnamefont {Mitarai}},\ and\ \bibinfo {author}
  {\bibfnamefont {M.}~\bibnamefont {Kitagawa}},\ }\bibfield  {title} {\bibinfo
  {title} {Boosting computational power through spatial multiplexing in quantum
  reservoir computing},\ }\href@noop {} {\bibfield  {journal} {\bibinfo
  {journal} {Physical Review Applied}\ }\textbf {\bibinfo {volume} {11}},\
  \bibinfo {pages} {034021} (\bibinfo {year} {2019})}\BibitemShut {NoStop}%
\bibitem [{\citenamefont {Chen}\ and\ \citenamefont
  {Nurdin}(2019)}]{chen2019learning}%
  \BibitemOpen
  \bibfield  {author} {\bibinfo {author} {\bibfnamefont {J.}~\bibnamefont
  {Chen}}\ and\ \bibinfo {author} {\bibfnamefont {H.~I.}\ \bibnamefont
  {Nurdin}},\ }\bibfield  {title} {\bibinfo {title} {Learning nonlinear
  input--output maps with dissipative quantum systems},\ }\href@noop {}
  {\bibfield  {journal} {\bibinfo  {journal} {Quantum Information Processing}\
  }\textbf {\bibinfo {volume} {18}},\ \bibinfo {pages} {198} (\bibinfo {year}
  {2019})}\BibitemShut {NoStop}%
\bibitem [{\citenamefont {Tran}\ and\ \citenamefont
  {Nakajima}(2020)}]{tran2020higher}%
  \BibitemOpen
  \bibfield  {author} {\bibinfo {author} {\bibfnamefont {Q.~H.}\ \bibnamefont
  {Tran}}\ and\ \bibinfo {author} {\bibfnamefont {K.}~\bibnamefont
  {Nakajima}},\ }\bibfield  {title} {\bibinfo {title} {Higher-order quantum
  reservoir computing},\ }\href@noop {} {\bibfield  {journal} {\bibinfo
  {journal} {arXiv preprint arXiv:2006.08999}\ } (\bibinfo {year}
  {2020})}\BibitemShut {NoStop}%
\bibitem [{\citenamefont {Mart{\'\i}nez-Pe{\~n}a}\ \emph
  {et~al.}(2020)\citenamefont {Mart{\'\i}nez-Pe{\~n}a}, \citenamefont
  {Nokkala}, \citenamefont {Giorgi}, \citenamefont {Zambrini},\ and\
  \citenamefont {Soriano}}]{martinez2020information}%
  \BibitemOpen
  \bibfield  {author} {\bibinfo {author} {\bibfnamefont {R.}~\bibnamefont
  {Mart{\'\i}nez-Pe{\~n}a}}, \bibinfo {author} {\bibfnamefont {J.}~\bibnamefont
  {Nokkala}}, \bibinfo {author} {\bibfnamefont {G.~L.}\ \bibnamefont {Giorgi}},
  \bibinfo {author} {\bibfnamefont {R.}~\bibnamefont {Zambrini}},\ and\
  \bibinfo {author} {\bibfnamefont {M.~C.}\ \bibnamefont {Soriano}},\
  }\bibfield  {title} {\bibinfo {title} {Information processing capacity of
  spin-based quantum reservoir computing systems},\ }\href@noop {} {\bibfield
  {journal} {\bibinfo  {journal} {Cognitive Computation}\ ,\ \bibinfo {pages}
  {1}} (\bibinfo {year} {2020})}\BibitemShut {NoStop}%
\bibitem [{\citenamefont {Ghosh}\ \emph
  {et~al.}(2019{\natexlab{a}})\citenamefont {Ghosh}, \citenamefont {Opala},
  \citenamefont {Matuszewski}, \citenamefont {Paterek},\ and\ \citenamefont
  {Liew}}]{ghosh2019quantum}%
  \BibitemOpen
  \bibfield  {author} {\bibinfo {author} {\bibfnamefont {S.}~\bibnamefont
  {Ghosh}}, \bibinfo {author} {\bibfnamefont {A.}~\bibnamefont {Opala}},
  \bibinfo {author} {\bibfnamefont {M.}~\bibnamefont {Matuszewski}}, \bibinfo
  {author} {\bibfnamefont {T.}~\bibnamefont {Paterek}},\ and\ \bibinfo {author}
  {\bibfnamefont {T.~C.}\ \bibnamefont {Liew}},\ }\bibfield  {title} {\bibinfo
  {title} {Quantum reservoir processing},\ }\href@noop {} {\bibfield  {journal}
  {\bibinfo  {journal} {npj Quantum Information}\ }\textbf {\bibinfo {volume}
  {5}},\ \bibinfo {pages} {35} (\bibinfo {year}
  {2019}{\natexlab{a}})}\BibitemShut {NoStop}%
\bibitem [{\citenamefont {Ghosh}\ \emph
  {et~al.}(2019{\natexlab{b}})\citenamefont {Ghosh}, \citenamefont {Paterek},\
  and\ \citenamefont {Liew}}]{ghosh2019neuromorphic}%
  \BibitemOpen
  \bibfield  {author} {\bibinfo {author} {\bibfnamefont {S.}~\bibnamefont
  {Ghosh}}, \bibinfo {author} {\bibfnamefont {T.}~\bibnamefont {Paterek}},\
  and\ \bibinfo {author} {\bibfnamefont {T.~C.~H.}\ \bibnamefont {Liew}},\
  }\bibfield  {title} {\bibinfo {title} {Quantum neuromorphic platform for
  quantum state preparation},\ }\href
  {https://doi.org/10.1103/PhysRevLett.123.260404} {\bibfield  {journal}
  {\bibinfo  {journal} {Phys. Rev. Lett.}\ }\textbf {\bibinfo {volume} {123}},\
  \bibinfo {pages} {260404} (\bibinfo {year} {2019}{\natexlab{b}})}\BibitemShut
  {NoStop}%
\bibitem [{\citenamefont {Nokkala}\ \emph {et~al.}(2021)\citenamefont
  {Nokkala}, \citenamefont {Mart{\'i}nez-Pe{\~{n}}a}, \citenamefont {Giorgi},
  \citenamefont {Parigi}, \citenamefont {Soriano},\ and\ \citenamefont
  {Zambrini}}]{nokkala2020gaussian}%
  \BibitemOpen
  \bibfield  {author} {\bibinfo {author} {\bibfnamefont {J.}~\bibnamefont
  {Nokkala}}, \bibinfo {author} {\bibfnamefont {R.}~\bibnamefont
  {Mart{\'i}nez-Pe{\~{n}}a}}, \bibinfo {author} {\bibfnamefont {G.~L.}\
  \bibnamefont {Giorgi}}, \bibinfo {author} {\bibfnamefont {V.}~\bibnamefont
  {Parigi}}, \bibinfo {author} {\bibfnamefont {M.~C.}\ \bibnamefont
  {Soriano}},\ and\ \bibinfo {author} {\bibfnamefont {R.}~\bibnamefont
  {Zambrini}},\ }\bibfield  {title} {\bibinfo {title} {Gaussian states of
  continuous-variable quantum systems provide universal and versatile reservoir
  computing},\ }\href {https://doi.org/10.1038/s42005-021-00556-w} {\bibfield
  {journal} {\bibinfo  {journal} {Communications Physics}\ }\textbf {\bibinfo
  {volume} {4}},\ \bibinfo {pages} {53} (\bibinfo {year} {2021})}\BibitemShut
  {NoStop}%
\bibitem [{\citenamefont {Govia}\ \emph {et~al.}(2021)\citenamefont {Govia},
  \citenamefont {Ribeill}, \citenamefont {Rowlands}, \citenamefont {Krovi},\
  and\ \citenamefont {Ohki}}]{govia2020quantum}%
  \BibitemOpen
  \bibfield  {author} {\bibinfo {author} {\bibfnamefont {L.~C.~G.}\
  \bibnamefont {Govia}}, \bibinfo {author} {\bibfnamefont {G.~J.}\ \bibnamefont
  {Ribeill}}, \bibinfo {author} {\bibfnamefont {G.~E.}\ \bibnamefont
  {Rowlands}}, \bibinfo {author} {\bibfnamefont {H.~K.}\ \bibnamefont
  {Krovi}},\ and\ \bibinfo {author} {\bibfnamefont {T.~A.}\ \bibnamefont
  {Ohki}},\ }\bibfield  {title} {\bibinfo {title} {Quantum reservoir computing
  with a single nonlinear oscillator},\ }\href
  {https://doi.org/10.1103/PhysRevResearch.3.013077} {\bibfield  {journal}
  {\bibinfo  {journal} {Phys. Rev. Research}\ }\textbf {\bibinfo {volume}
  {3}},\ \bibinfo {pages} {013077} (\bibinfo {year} {2021})}\BibitemShut
  {NoStop}%
\bibitem [{\citenamefont {Ghosh}\ \emph {et~al.}(2021)\citenamefont {Ghosh},
  \citenamefont {Krisnanda}, \citenamefont {Paterek},\ and\ \citenamefont
  {Liew}}]{ghosh2020universal}%
  \BibitemOpen
  \bibfield  {author} {\bibinfo {author} {\bibfnamefont {S.}~\bibnamefont
  {Ghosh}}, \bibinfo {author} {\bibfnamefont {T.}~\bibnamefont {Krisnanda}},
  \bibinfo {author} {\bibfnamefont {T.}~\bibnamefont {Paterek}},\ and\ \bibinfo
  {author} {\bibfnamefont {T.~C.~H.}\ \bibnamefont {Liew}},\ }\bibfield
  {title} {\bibinfo {title} {Realising and compressing quantum circuits with
  quantum reservoir computing},\ }\href
  {https://doi.org/10.1038/s42005-021-00606-3} {\bibfield  {journal} {\bibinfo
  {journal} {Communications Physics}\ }\textbf {\bibinfo {volume} {4}},\
  \bibinfo {pages} {105} (\bibinfo {year} {2021})}\BibitemShut {NoStop}%
\bibitem [{\citenamefont {Negoro}\ \emph {et~al.}(2018)\citenamefont {Negoro},
  \citenamefont {Mitarai}, \citenamefont {Fujii}, \citenamefont {Nakajima},\
  and\ \citenamefont {Kitagawa}}]{negoro2018machine}%
  \BibitemOpen
  \bibfield  {author} {\bibinfo {author} {\bibfnamefont {M.}~\bibnamefont
  {Negoro}}, \bibinfo {author} {\bibfnamefont {K.}~\bibnamefont {Mitarai}},
  \bibinfo {author} {\bibfnamefont {K.}~\bibnamefont {Fujii}}, \bibinfo
  {author} {\bibfnamefont {K.}~\bibnamefont {Nakajima}},\ and\ \bibinfo
  {author} {\bibfnamefont {M.}~\bibnamefont {Kitagawa}},\ }\bibfield  {title}
  {\bibinfo {title} {Machine learning with controllable quantum dynamics of a
  nuclear spin ensemble in a solid},\ }\href@noop {} {\bibfield  {journal}
  {\bibinfo  {journal} {arXiv preprint arXiv:1806.10910}\ } (\bibinfo {year}
  {2018})}\BibitemShut {NoStop}%
\bibitem [{\citenamefont {Chen}\ \emph {et~al.}(2020)\citenamefont {Chen},
  \citenamefont {Nurdin},\ and\ \citenamefont {Yamamoto}}]{chen2020temporal}%
  \BibitemOpen
  \bibfield  {author} {\bibinfo {author} {\bibfnamefont {J.}~\bibnamefont
  {Chen}}, \bibinfo {author} {\bibfnamefont {H.~I.}\ \bibnamefont {Nurdin}},\
  and\ \bibinfo {author} {\bibfnamefont {N.}~\bibnamefont {Yamamoto}},\
  }\bibfield  {title} {\bibinfo {title} {Temporal information processing on
  noisy quantum computers},\ }\href
  {https://doi.org/10.1103/PhysRevApplied.14.024065} {\bibfield  {journal}
  {\bibinfo  {journal} {Phys. Rev. Applied}\ }\textbf {\bibinfo {volume}
  {14}},\ \bibinfo {pages} {024065} (\bibinfo {year} {2020})}\BibitemShut
  {NoStop}%
\bibitem [{\citenamefont {Deutsch}(1991)}]{deutsch1991quantum}%
  \BibitemOpen
  \bibfield  {author} {\bibinfo {author} {\bibfnamefont {J.~M.}\ \bibnamefont
  {Deutsch}},\ }\bibfield  {title} {\bibinfo {title} {Quantum statistical
  mechanics in a closed system},\ }\href@noop {} {\bibfield  {journal}
  {\bibinfo  {journal} {Physical Review A}\ }\textbf {\bibinfo {volume} {43}},\
  \bibinfo {pages} {2046} (\bibinfo {year} {1991})}\BibitemShut {NoStop}%
\bibitem [{\citenamefont {Srednicki}(1999)}]{srednicki1999approach}%
  \BibitemOpen
  \bibfield  {author} {\bibinfo {author} {\bibfnamefont {M.}~\bibnamefont
  {Srednicki}},\ }\bibfield  {title} {\bibinfo {title} {The approach to thermal
  equilibrium in quantized chaotic systems},\ }\href@noop {} {\bibfield
  {journal} {\bibinfo  {journal} {Journal of Physics A: Mathematical and
  General}\ }\textbf {\bibinfo {volume} {32}},\ \bibinfo {pages} {1163}
  (\bibinfo {year} {1999})}\BibitemShut {NoStop}%
\bibitem [{\citenamefont {D'Alessio}\ \emph {et~al.}(2016)\citenamefont
  {D'Alessio}, \citenamefont {Kafri}, \citenamefont {Polkovnikov},\ and\
  \citenamefont {Rigol}}]{d2016quantum}%
  \BibitemOpen
  \bibfield  {author} {\bibinfo {author} {\bibfnamefont {L.}~\bibnamefont
  {D'Alessio}}, \bibinfo {author} {\bibfnamefont {Y.}~\bibnamefont {Kafri}},
  \bibinfo {author} {\bibfnamefont {A.}~\bibnamefont {Polkovnikov}},\ and\
  \bibinfo {author} {\bibfnamefont {M.}~\bibnamefont {Rigol}},\ }\bibfield
  {title} {\bibinfo {title} {From quantum chaos and eigenstate thermalization
  to statistical mechanics and thermodynamics},\ }\href@noop {} {\bibfield
  {journal} {\bibinfo  {journal} {Advances in Physics}\ }\textbf {\bibinfo
  {volume} {65}},\ \bibinfo {pages} {239} (\bibinfo {year} {2016})}\BibitemShut
  {NoStop}%
\bibitem [{\citenamefont {Abanin}\ \emph {et~al.}(2019)\citenamefont {Abanin},
  \citenamefont {Altman}, \citenamefont {Bloch},\ and\ \citenamefont
  {Serbyn}}]{abanin2019colloquium}%
  \BibitemOpen
  \bibfield  {author} {\bibinfo {author} {\bibfnamefont {D.~A.}\ \bibnamefont
  {Abanin}}, \bibinfo {author} {\bibfnamefont {E.}~\bibnamefont {Altman}},
  \bibinfo {author} {\bibfnamefont {I.}~\bibnamefont {Bloch}},\ and\ \bibinfo
  {author} {\bibfnamefont {M.}~\bibnamefont {Serbyn}},\ }\bibfield  {title}
  {\bibinfo {title} {Colloquium: Many-body localization, thermalization, and
  entanglement},\ }\href@noop {} {\bibfield  {journal} {\bibinfo  {journal}
  {Reviews of Modern Physics}\ }\textbf {\bibinfo {volume} {91}},\ \bibinfo
  {pages} {021001} (\bibinfo {year} {2019})}\BibitemShut {NoStop}%
\bibitem [{\citenamefont {{\v{Z}}unkovi{\v{c}}}\ \emph
  {et~al.}(2018)\citenamefont {{\v{Z}}unkovi{\v{c}}}, \citenamefont {Heyl},
  \citenamefont {Knap},\ and\ \citenamefont {Silva}}]{vzunkovivc2018dynamical}%
  \BibitemOpen
  \bibfield  {author} {\bibinfo {author} {\bibfnamefont {B.}~\bibnamefont
  {{\v{Z}}unkovi{\v{c}}}}, \bibinfo {author} {\bibfnamefont {M.}~\bibnamefont
  {Heyl}}, \bibinfo {author} {\bibfnamefont {M.}~\bibnamefont {Knap}},\ and\
  \bibinfo {author} {\bibfnamefont {A.}~\bibnamefont {Silva}},\ }\bibfield
  {title} {\bibinfo {title} {Dynamical quantum phase transitions in spin chains
  with long-range interactions: Merging different concepts of nonequilibrium
  criticality},\ }\href@noop {} {\bibfield  {journal} {\bibinfo  {journal}
  {Physical review letters}\ }\textbf {\bibinfo {volume} {120}},\ \bibinfo
  {pages} {130601} (\bibinfo {year} {2018})}\BibitemShut {NoStop}%
\bibitem [{\citenamefont {Huse}\ \emph {et~al.}(2013)\citenamefont {Huse},
  \citenamefont {Nandkishore}, \citenamefont {Oganesyan}, \citenamefont {Pal},\
  and\ \citenamefont {Sondhi}}]{huse2013localization}%
  \BibitemOpen
  \bibfield  {author} {\bibinfo {author} {\bibfnamefont {D.~A.}\ \bibnamefont
  {Huse}}, \bibinfo {author} {\bibfnamefont {R.}~\bibnamefont {Nandkishore}},
  \bibinfo {author} {\bibfnamefont {V.}~\bibnamefont {Oganesyan}}, \bibinfo
  {author} {\bibfnamefont {A.}~\bibnamefont {Pal}},\ and\ \bibinfo {author}
  {\bibfnamefont {S.~L.}\ \bibnamefont {Sondhi}},\ }\bibfield  {title}
  {\bibinfo {title} {Localization-protected quantum order},\ }\href@noop {}
  {\bibfield  {journal} {\bibinfo  {journal} {Physical Review B}\ }\textbf
  {\bibinfo {volume} {88}},\ \bibinfo {pages} {014206} (\bibinfo {year}
  {2013})}\BibitemShut {NoStop}%
\bibitem [{\citenamefont {Ponte}\ \emph {et~al.}(2015)\citenamefont {Ponte},
  \citenamefont {Chandran}, \citenamefont {Papi{\'c}},\ and\ \citenamefont
  {Abanin}}]{ponte2015periodically}%
  \BibitemOpen
  \bibfield  {author} {\bibinfo {author} {\bibfnamefont {P.}~\bibnamefont
  {Ponte}}, \bibinfo {author} {\bibfnamefont {A.}~\bibnamefont {Chandran}},
  \bibinfo {author} {\bibfnamefont {Z.}~\bibnamefont {Papi{\'c}}},\ and\
  \bibinfo {author} {\bibfnamefont {D.~A.}\ \bibnamefont {Abanin}},\ }\bibfield
   {title} {\bibinfo {title} {Periodically driven ergodic and many-body
  localized quantum systems},\ }\href@noop {} {\bibfield  {journal} {\bibinfo
  {journal} {Annals of Physics}\ }\textbf {\bibinfo {volume} {353}},\ \bibinfo
  {pages} {196} (\bibinfo {year} {2015})}\BibitemShut {NoStop}%
\bibitem [{\citenamefont {Lazarides}\ \emph {et~al.}(2015)\citenamefont
  {Lazarides}, \citenamefont {Das},\ and\ \citenamefont
  {Moessner}}]{lazarides2015fate}%
  \BibitemOpen
  \bibfield  {author} {\bibinfo {author} {\bibfnamefont {A.}~\bibnamefont
  {Lazarides}}, \bibinfo {author} {\bibfnamefont {A.}~\bibnamefont {Das}},\
  and\ \bibinfo {author} {\bibfnamefont {R.}~\bibnamefont {Moessner}},\
  }\bibfield  {title} {\bibinfo {title} {Fate of many-body localization under
  periodic driving},\ }\href@noop {} {\bibfield  {journal} {\bibinfo  {journal}
  {Physical review letters}\ }\textbf {\bibinfo {volume} {115}},\ \bibinfo
  {pages} {030402} (\bibinfo {year} {2015})}\BibitemShut {NoStop}%
\bibitem [{\citenamefont {Tangpanitanon}\ \emph {et~al.}(2020)\citenamefont
  {Tangpanitanon}, \citenamefont {Thanasilp}, \citenamefont {Dangniam},
  \citenamefont {Lemonde},\ and\ \citenamefont
  {Angelakis}}]{tangpanitanon2020expressibility}%
  \BibitemOpen
  \bibfield  {author} {\bibinfo {author} {\bibfnamefont {J.}~\bibnamefont
  {Tangpanitanon}}, \bibinfo {author} {\bibfnamefont {S.}~\bibnamefont
  {Thanasilp}}, \bibinfo {author} {\bibfnamefont {N.}~\bibnamefont {Dangniam}},
  \bibinfo {author} {\bibfnamefont {M.-A.}\ \bibnamefont {Lemonde}},\ and\
  \bibinfo {author} {\bibfnamefont {D.~G.}\ \bibnamefont {Angelakis}},\
  }\bibfield  {title} {\bibinfo {title} {Expressibility and trainability of
  parametrized analog quantum systems for machine learning applications},\
  }\href {https://doi.org/10.1103/PhysRevResearch.2.043364} {\bibfield
  {journal} {\bibinfo  {journal} {Phys. Rev. Research}\ }\textbf {\bibinfo
  {volume} {2}},\ \bibinfo {pages} {043364} (\bibinfo {year}
  {2020})}\BibitemShut {NoStop}%
\bibitem [{\citenamefont {Altshuler}\ \emph {et~al.}(2010)\citenamefont
  {Altshuler}, \citenamefont {Krovi},\ and\ \citenamefont
  {Roland}}]{altshuler2010anderson}%
  \BibitemOpen
  \bibfield  {author} {\bibinfo {author} {\bibfnamefont {B.}~\bibnamefont
  {Altshuler}}, \bibinfo {author} {\bibfnamefont {H.}~\bibnamefont {Krovi}},\
  and\ \bibinfo {author} {\bibfnamefont {J.}~\bibnamefont {Roland}},\
  }\bibfield  {title} {\bibinfo {title} {Anderson localization makes adiabatic
  quantum optimization fail},\ }\href@noop {} {\bibfield  {journal} {\bibinfo
  {journal} {Proceedings of the National Academy of Sciences}\ }\textbf
  {\bibinfo {volume} {107}},\ \bibinfo {pages} {12446} (\bibinfo {year}
  {2010})}\BibitemShut {NoStop}%
\bibitem [{\citenamefont {Laumann}\ \emph {et~al.}(2015)\citenamefont
  {Laumann}, \citenamefont {Moessner}, \citenamefont {Scardicchio},\ and\
  \citenamefont {Sondhi}}]{laumann2015quantum}%
  \BibitemOpen
  \bibfield  {author} {\bibinfo {author} {\bibfnamefont {C.~R.}\ \bibnamefont
  {Laumann}}, \bibinfo {author} {\bibfnamefont {R.}~\bibnamefont {Moessner}},
  \bibinfo {author} {\bibfnamefont {A.}~\bibnamefont {Scardicchio}},\ and\
  \bibinfo {author} {\bibfnamefont {S.~L.}\ \bibnamefont {Sondhi}},\ }\bibfield
   {title} {\bibinfo {title} {Quantum annealing: The fastest route to quantum
  computation?},\ }\href@noop {} {\bibfield  {journal} {\bibinfo  {journal}
  {The European Physical Journal Special Topics}\ }\textbf {\bibinfo {volume}
  {224}},\ \bibinfo {pages} {75} (\bibinfo {year} {2015})}\BibitemShut
  {NoStop}%
\bibitem [{\citenamefont {Keating}\ \emph {et~al.}(2007)\citenamefont
  {Keating}, \citenamefont {Linden}, \citenamefont {Matthews},\ and\
  \citenamefont {Winter}}]{keating2007localization}%
  \BibitemOpen
  \bibfield  {author} {\bibinfo {author} {\bibfnamefont {J.}~\bibnamefont
  {Keating}}, \bibinfo {author} {\bibfnamefont {N.}~\bibnamefont {Linden}},
  \bibinfo {author} {\bibfnamefont {J.}~\bibnamefont {Matthews}},\ and\
  \bibinfo {author} {\bibfnamefont {A.}~\bibnamefont {Winter}},\ }\bibfield
  {title} {\bibinfo {title} {Localization and its consequences for quantum walk
  algorithms and quantum communication},\ }\href@noop {} {\bibfield  {journal}
  {\bibinfo  {journal} {Physical Review A}\ }\textbf {\bibinfo {volume} {76}},\
  \bibinfo {pages} {012315} (\bibinfo {year} {2007})}\BibitemShut {NoStop}%
\bibitem [{\citenamefont {Schreiber}\ \emph {et~al.}(2011)\citenamefont
  {Schreiber}, \citenamefont {Cassemiro}, \citenamefont {Poto{\v{c}}ek},
  \citenamefont {G{\'a}bris}, \citenamefont {Jex},\ and\ \citenamefont
  {Silberhorn}}]{schreiber2011decoherence}%
  \BibitemOpen
  \bibfield  {author} {\bibinfo {author} {\bibfnamefont {A.}~\bibnamefont
  {Schreiber}}, \bibinfo {author} {\bibfnamefont {K.}~\bibnamefont
  {Cassemiro}}, \bibinfo {author} {\bibfnamefont {V.}~\bibnamefont
  {Poto{\v{c}}ek}}, \bibinfo {author} {\bibfnamefont {A.}~\bibnamefont
  {G{\'a}bris}}, \bibinfo {author} {\bibfnamefont {I.}~\bibnamefont {Jex}},\
  and\ \bibinfo {author} {\bibfnamefont {C.}~\bibnamefont {Silberhorn}},\
  }\bibfield  {title} {\bibinfo {title} {Decoherence and disorder in quantum
  walks: from ballistic spread to localization},\ }\href@noop {} {\bibfield
  {journal} {\bibinfo  {journal} {Physical review letters}\ }\textbf {\bibinfo
  {volume} {106}},\ \bibinfo {pages} {180403} (\bibinfo {year}
  {2011})}\BibitemShut {NoStop}%
\bibitem [{\citenamefont {Oganesyan}\ and\ \citenamefont
  {Huse}(2007)}]{oganesyan2007localization}%
  \BibitemOpen
  \bibfield  {author} {\bibinfo {author} {\bibfnamefont {V.}~\bibnamefont
  {Oganesyan}}\ and\ \bibinfo {author} {\bibfnamefont {D.~A.}\ \bibnamefont
  {Huse}},\ }\bibfield  {title} {\bibinfo {title} {Localization of interacting
  fermions at high temperature},\ }\href@noop {} {\bibfield  {journal}
  {\bibinfo  {journal} {Physical Review B}\ }\textbf {\bibinfo {volume} {75}},\
  \bibinfo {pages} {155111} (\bibinfo {year} {2007})}\BibitemShut {NoStop}%
\bibitem [{sup()}]{suppmat}%
  \BibitemOpen
  \href@noop {} {\bibinfo {title} {See {Supplemental} {Material} for additional
  notes on the dynamics, the convergence properties of the spin network, a note
  on the experimental implementations, the {NARMA} task, and the {IPC} of the
  system, including {R}efs. \cite{ weinberg2016quspin,johansson2012qutip,
  ilievski2015complete,vosk2014dynamical,
  geraedts2017emergent,zobov2017effect,wei2018exploring,richerme2014non,huang2020predicting,kubota2019dynamical}}}\BibitemShut
  {NoStop}%
\bibitem [{\citenamefont {Atas}\ \emph {et~al.}(2013)\citenamefont {Atas},
  \citenamefont {Bogomolny}, \citenamefont {Giraud},\ and\ \citenamefont
  {Roux}}]{atas2013distribution}%
  \BibitemOpen
  \bibfield  {author} {\bibinfo {author} {\bibfnamefont {Y.}~\bibnamefont
  {Atas}}, \bibinfo {author} {\bibfnamefont {E.}~\bibnamefont {Bogomolny}},
  \bibinfo {author} {\bibfnamefont {O.}~\bibnamefont {Giraud}},\ and\ \bibinfo
  {author} {\bibfnamefont {G.}~\bibnamefont {Roux}},\ }\bibfield  {title}
  {\bibinfo {title} {Distribution of the ratio of consecutive level spacings in
  random matrix ensembles},\ }\href@noop {} {\bibfield  {journal} {\bibinfo
  {journal} {Physical Review Letters}\ }\textbf {\bibinfo {volume} {110}},\
  \bibinfo {pages} {084101} (\bibinfo {year} {2013})}\BibitemShut {NoStop}%
\bibitem [{\citenamefont {Baldwin}\ \emph {et~al.}(2017)\citenamefont
  {Baldwin}, \citenamefont {Laumann}, \citenamefont {Pal},\ and\ \citenamefont
  {Scardicchio}}]{baldwin2017clustering}%
  \BibitemOpen
  \bibfield  {author} {\bibinfo {author} {\bibfnamefont {C.}~\bibnamefont
  {Baldwin}}, \bibinfo {author} {\bibfnamefont {C.}~\bibnamefont {Laumann}},
  \bibinfo {author} {\bibfnamefont {A.}~\bibnamefont {Pal}},\ and\ \bibinfo
  {author} {\bibfnamefont {A.}~\bibnamefont {Scardicchio}},\ }\bibfield
  {title} {\bibinfo {title} {Clustering of nonergodic eigenstates in quantum
  spin glasses},\ }\href@noop {} {\bibfield  {journal} {\bibinfo  {journal}
  {Physical Review Letters}\ }\textbf {\bibinfo {volume} {118}},\ \bibinfo
  {pages} {127201} (\bibinfo {year} {2017})}\BibitemShut {NoStop}%
\bibitem [{\citenamefont {Mukherjee}\ \emph {et~al.}(2018)\citenamefont
  {Mukherjee}, \citenamefont {Nag},\ and\ \citenamefont
  {Garg}}]{mukherjee2018many}%
  \BibitemOpen
  \bibfield  {author} {\bibinfo {author} {\bibfnamefont {S.}~\bibnamefont
  {Mukherjee}}, \bibinfo {author} {\bibfnamefont {S.}~\bibnamefont {Nag}},\
  and\ \bibinfo {author} {\bibfnamefont {A.}~\bibnamefont {Garg}},\ }\bibfield
  {title} {\bibinfo {title} {Many-body localization-delocalization transition
  in the quantum sherrington-kirkpatrick model},\ }\href@noop {} {\bibfield
  {journal} {\bibinfo  {journal} {Physical Review B}\ }\textbf {\bibinfo
  {volume} {97}},\ \bibinfo {pages} {144202} (\bibinfo {year}
  {2018})}\BibitemShut {NoStop}%
\bibitem [{\citenamefont {Rademaker}\ and\ \citenamefont
  {Abanin}(2020)}]{rademaker2019bridging}%
  \BibitemOpen
  \bibfield  {author} {\bibinfo {author} {\bibfnamefont {L.}~\bibnamefont
  {Rademaker}}\ and\ \bibinfo {author} {\bibfnamefont {D.~A.}\ \bibnamefont
  {Abanin}},\ }\bibfield  {title} {\bibinfo {title} {Slow nonthermalizing
  dynamics in a quantum spin glass},\ }\href
  {https://doi.org/10.1103/PhysRevLett.125.260405} {\bibfield  {journal}
  {\bibinfo  {journal} {Phys. Rev. Lett.}\ }\textbf {\bibinfo {volume} {125}},\
  \bibinfo {pages} {260405} (\bibinfo {year} {2020})}\BibitemShut {NoStop}%
\bibitem [{\citenamefont {Maksymov}\ and\ \citenamefont
  {Burin}(2020)}]{maksymov2020many}%
  \BibitemOpen
  \bibfield  {author} {\bibinfo {author} {\bibfnamefont {A.~O.}\ \bibnamefont
  {Maksymov}}\ and\ \bibinfo {author} {\bibfnamefont {A.~L.}\ \bibnamefont
  {Burin}},\ }\bibfield  {title} {\bibinfo {title} {Many-body localization in
  spin chains with long-range transverse interactions: Scaling of critical
  disorder with system size},\ }\href@noop {} {\bibfield  {journal} {\bibinfo
  {journal} {Physical Review B}\ }\textbf {\bibinfo {volume} {101}},\ \bibinfo
  {pages} {024201} (\bibinfo {year} {2020})}\BibitemShut {NoStop}%
\bibitem [{\citenamefont {Smith}\ \emph {et~al.}(2016)\citenamefont {Smith},
  \citenamefont {Lee}, \citenamefont {Richerme}, \citenamefont {Neyenhuis},
  \citenamefont {Hess}, \citenamefont {Hauke}, \citenamefont {Heyl},
  \citenamefont {Huse},\ and\ \citenamefont {Monroe}}]{smith2016many}%
  \BibitemOpen
  \bibfield  {author} {\bibinfo {author} {\bibfnamefont {J.}~\bibnamefont
  {Smith}}, \bibinfo {author} {\bibfnamefont {A.}~\bibnamefont {Lee}}, \bibinfo
  {author} {\bibfnamefont {P.}~\bibnamefont {Richerme}}, \bibinfo {author}
  {\bibfnamefont {B.}~\bibnamefont {Neyenhuis}}, \bibinfo {author}
  {\bibfnamefont {P.~W.}\ \bibnamefont {Hess}}, \bibinfo {author}
  {\bibfnamefont {P.}~\bibnamefont {Hauke}}, \bibinfo {author} {\bibfnamefont
  {M.}~\bibnamefont {Heyl}}, \bibinfo {author} {\bibfnamefont {D.~A.}\
  \bibnamefont {Huse}},\ and\ \bibinfo {author} {\bibfnamefont
  {C.}~\bibnamefont {Monroe}},\ }\bibfield  {title} {\bibinfo {title}
  {Many-body localization in a quantum simulator with programmable random
  disorder},\ }\href@noop {} {\bibfield  {journal} {\bibinfo  {journal} {Nature
  Physics}\ }\textbf {\bibinfo {volume} {12}},\ \bibinfo {pages} {907}
  (\bibinfo {year} {2016})}\BibitemShut {NoStop}%
\bibitem [{\citenamefont {Moudgalya}\ \emph {et~al.}(2020)\citenamefont
  {Moudgalya}, \citenamefont {Huse},\ and\ \citenamefont
  {Khemani}}]{moudgalya2020perturbative}%
  \BibitemOpen
  \bibfield  {author} {\bibinfo {author} {\bibfnamefont {S.}~\bibnamefont
  {Moudgalya}}, \bibinfo {author} {\bibfnamefont {D.~A.}\ \bibnamefont
  {Huse}},\ and\ \bibinfo {author} {\bibfnamefont {V.}~\bibnamefont
  {Khemani}},\ }\bibfield  {title} {\bibinfo {title} {Perturbative instability
  towards delocalization at phase transitions between {MBL} phases},\
  }\href@noop {} {\bibfield  {journal} {\bibinfo  {journal} {arXiv preprint
  arXiv:2008.09113}\ } (\bibinfo {year} {2020})}\BibitemShut {NoStop}%
\bibitem [{\citenamefont {Sahay}\ \emph {et~al.}(2021)\citenamefont {Sahay},
  \citenamefont {Machado}, \citenamefont {Ye}, \citenamefont {Laumann},\ and\
  \citenamefont {Yao}}]{sahay2020emergent}%
  \BibitemOpen
  \bibfield  {author} {\bibinfo {author} {\bibfnamefont {R.}~\bibnamefont
  {Sahay}}, \bibinfo {author} {\bibfnamefont {F.}~\bibnamefont {Machado}},
  \bibinfo {author} {\bibfnamefont {B.}~\bibnamefont {Ye}}, \bibinfo {author}
  {\bibfnamefont {C.~R.}\ \bibnamefont {Laumann}},\ and\ \bibinfo {author}
  {\bibfnamefont {N.~Y.}\ \bibnamefont {Yao}},\ }\bibfield  {title} {\bibinfo
  {title} {Emergent ergodicity at the transition between many-body localized
  phases},\ }\href {https://doi.org/10.1103/PhysRevLett.126.100604} {\bibfield
  {journal} {\bibinfo  {journal} {Phys. Rev. Lett.}\ }\textbf {\bibinfo
  {volume} {126}},\ \bibinfo {pages} {100604} (\bibinfo {year}
  {2021})}\BibitemShut {NoStop}%
\bibitem [{\citenamefont {Khemani}\ \emph {et~al.}(2017)\citenamefont
  {Khemani}, \citenamefont {Lim}, \citenamefont {Sheng},\ and\ \citenamefont
  {Huse}}]{khemani2017}%
  \BibitemOpen
  \bibfield  {author} {\bibinfo {author} {\bibfnamefont {V.}~\bibnamefont
  {Khemani}}, \bibinfo {author} {\bibfnamefont {S.~P.}\ \bibnamefont {Lim}},
  \bibinfo {author} {\bibfnamefont {D.~N.}\ \bibnamefont {Sheng}},\ and\
  \bibinfo {author} {\bibfnamefont {D.~A.}\ \bibnamefont {Huse}},\ }\bibfield
  {title} {\bibinfo {title} {Critical properties of the many-body localization
  transition},\ }\href {https://doi.org/10.1103/PhysRevX.7.021013} {\bibfield
  {journal} {\bibinfo  {journal} {Phys. Rev. X}\ }\textbf {\bibinfo {volume}
  {7}},\ \bibinfo {pages} {021013} (\bibinfo {year} {2017})}\BibitemShut
  {NoStop}%
\bibitem [{\citenamefont {Panda}\ \emph {et~al.}(2020)\citenamefont {Panda},
  \citenamefont {Scardicchio}, \citenamefont {Schulz}, \citenamefont {Taylor},\
  and\ \citenamefont {{\v{Z}}nidari{\v{c}}}}]{scardicchio_epl2020}%
  \BibitemOpen
  \bibfield  {author} {\bibinfo {author} {\bibfnamefont {R.~K.}\ \bibnamefont
  {Panda}}, \bibinfo {author} {\bibfnamefont {A.}~\bibnamefont {Scardicchio}},
  \bibinfo {author} {\bibfnamefont {M.}~\bibnamefont {Schulz}}, \bibinfo
  {author} {\bibfnamefont {S.~R.}\ \bibnamefont {Taylor}},\ and\ \bibinfo
  {author} {\bibfnamefont {M.}~\bibnamefont {{\v{Z}}nidari{\v{c}}}},\
  }\bibfield  {title} {\bibinfo {title} {Can we study the many-body
  localisation transition?},\ }\href
  {https://doi.org/10.1209/0295-5075/128/67003} {\bibfield  {journal} {\bibinfo
   {journal} {{EPL} (Europhysics Letters)}\ }\textbf {\bibinfo {volume}
  {128}},\ \bibinfo {pages} {67003} (\bibinfo {year} {2020})}\BibitemShut
  {NoStop}%
\bibitem [{\citenamefont {Grigoryeva}\ and\ \citenamefont
  {Ortega}(2018)}]{grigoryeva2018echo}%
  \BibitemOpen
  \bibfield  {author} {\bibinfo {author} {\bibfnamefont {L.}~\bibnamefont
  {Grigoryeva}}\ and\ \bibinfo {author} {\bibfnamefont {J.-P.}\ \bibnamefont
  {Ortega}},\ }\bibfield  {title} {\bibinfo {title} {Echo state networks are
  universal},\ }\href@noop {} {\bibfield  {journal} {\bibinfo  {journal}
  {Neural Networks}\ }\textbf {\bibinfo {volume} {108}},\ \bibinfo {pages}
  {495} (\bibinfo {year} {2018})}\BibitemShut {NoStop}%
\bibitem [{\citenamefont {Dambre}\ \emph {et~al.}(2012)\citenamefont {Dambre},
  \citenamefont {Verstraeten}, \citenamefont {Schrauwen},\ and\ \citenamefont
  {Massar}}]{dambre2012information}%
  \BibitemOpen
  \bibfield  {author} {\bibinfo {author} {\bibfnamefont {J.}~\bibnamefont
  {Dambre}}, \bibinfo {author} {\bibfnamefont {D.}~\bibnamefont {Verstraeten}},
  \bibinfo {author} {\bibfnamefont {B.}~\bibnamefont {Schrauwen}},\ and\
  \bibinfo {author} {\bibfnamefont {S.}~\bibnamefont {Massar}},\ }\bibfield
  {title} {\bibinfo {title} {Information processing capacity of dynamical
  systems},\ }\href@noop {} {\bibfield  {journal} {\bibinfo  {journal}
  {Scientific reports}\ }\textbf {\bibinfo {volume} {2}},\ \bibinfo {pages} {1}
  (\bibinfo {year} {2012})}\BibitemShut {NoStop}%
\bibitem [{\citenamefont {Atiya}\ and\ \citenamefont
  {Parlos}(2000)}]{atiya2000new}%
  \BibitemOpen
  \bibfield  {author} {\bibinfo {author} {\bibfnamefont {A.~F.}\ \bibnamefont
  {Atiya}}\ and\ \bibinfo {author} {\bibfnamefont {A.~G.}\ \bibnamefont
  {Parlos}},\ }\bibfield  {title} {\bibinfo {title} {New results on recurrent
  network training: unifying the algorithms and accelerating convergence},\
  }\href@noop {} {\bibfield  {journal} {\bibinfo  {journal} {IEEE transactions
  on neural networks}\ }\textbf {\bibinfo {volume} {11}},\ \bibinfo {pages}
  {697} (\bibinfo {year} {2000})}\BibitemShut {NoStop}%
\bibitem [{\citenamefont {Carroll}(2020)}]{carroll2020reservoir}%
  \BibitemOpen
  \bibfield  {author} {\bibinfo {author} {\bibfnamefont {T.~L.}\ \bibnamefont
  {Carroll}},\ }\bibfield  {title} {\bibinfo {title} {Do reservoir computers
  work best at the edge of chaos?},\ }\href@noop {} {\bibfield  {journal}
  {\bibinfo  {journal} {Chaos}\ }\textbf {\bibinfo {volume} {30}},\ \bibinfo
  {pages} {121109} (\bibinfo {year} {2020})}\BibitemShut {NoStop}%
\bibitem [{lac()}]{lackconv}%
  \BibitemOpen
  \href@noop {} {\bibinfo {title} {The localized regime shows an extremely much
  slower convergence than in the ergodic phase. {F}or practical purposes, this
  is definitely an obstacle for implementing {QRC} even if strictly speaking
  convergence would asymptotically be achieved since the map presented in {E}q.
  (2) of the manuscript is contractive due to the partial trace.}}\BibitemShut
  {Stop}%
\bibitem [{\citenamefont {Swingle}(2018)}]{swingle2018unscrambling}%
  \BibitemOpen
  \bibfield  {author} {\bibinfo {author} {\bibfnamefont {B.}~\bibnamefont
  {Swingle}},\ }\bibfield  {title} {\bibinfo {title} {Unscrambling the physics
  of out-of-time-order correlators},\ }\href@noop {} {\bibfield  {journal}
  {\bibinfo  {journal} {Nature Physics}\ }\textbf {\bibinfo {volume} {14}},\
  \bibinfo {pages} {988} (\bibinfo {year} {2018})}\BibitemShut {NoStop}%
\bibitem [{\citenamefont {Heyl}\ \emph {et~al.}(2013)\citenamefont {Heyl},
  \citenamefont {Polkovnikov},\ and\ \citenamefont {Kehrein}}]{heyl2013}%
  \BibitemOpen
  \bibfield  {author} {\bibinfo {author} {\bibfnamefont {M.}~\bibnamefont
  {Heyl}}, \bibinfo {author} {\bibfnamefont {A.}~\bibnamefont {Polkovnikov}},\
  and\ \bibinfo {author} {\bibfnamefont {S.}~\bibnamefont {Kehrein}},\
  }\bibfield  {title} {\bibinfo {title} {Dynamical quantum phase transitions in
  the transverse-field ising model},\ }\href
  {https://doi.org/10.1103/PhysRevLett.110.135704} {\bibfield  {journal}
  {\bibinfo  {journal} {Phys. Rev. Lett.}\ }\textbf {\bibinfo {volume} {110}},\
  \bibinfo {pages} {135704} (\bibinfo {year} {2013})}\BibitemShut {NoStop}%
\bibitem [{\citenamefont {Eddins}\ \emph {et~al.}(2019)\citenamefont {Eddins},
  \citenamefont {Kreikebaum}, \citenamefont {Toyli}, \citenamefont
  {Levenson-Falk}, \citenamefont {Dove}, \citenamefont {Livingston},
  \citenamefont {Levitan}, \citenamefont {Govia}, \citenamefont {Clerk},\ and\
  \citenamefont {Siddiqi}}]{eddins2019high}%
  \BibitemOpen
  \bibfield  {author} {\bibinfo {author} {\bibfnamefont {A.}~\bibnamefont
  {Eddins}}, \bibinfo {author} {\bibfnamefont {J.}~\bibnamefont {Kreikebaum}},
  \bibinfo {author} {\bibfnamefont {D.}~\bibnamefont {Toyli}}, \bibinfo
  {author} {\bibfnamefont {E.}~\bibnamefont {Levenson-Falk}}, \bibinfo {author}
  {\bibfnamefont {A.}~\bibnamefont {Dove}}, \bibinfo {author} {\bibfnamefont
  {W.}~\bibnamefont {Livingston}}, \bibinfo {author} {\bibfnamefont
  {B.}~\bibnamefont {Levitan}}, \bibinfo {author} {\bibfnamefont
  {L.}~\bibnamefont {Govia}}, \bibinfo {author} {\bibfnamefont
  {A.}~\bibnamefont {Clerk}},\ and\ \bibinfo {author} {\bibfnamefont
  {I.}~\bibnamefont {Siddiqi}},\ }\bibfield  {title} {\bibinfo {title}
  {High-efficiency measurement of an artificial atom embedded in a parametric
  amplifier},\ }\href@noop {} {\bibfield  {journal} {\bibinfo  {journal}
  {Physical Review X}\ }\textbf {\bibinfo {volume} {9}},\ \bibinfo {pages}
  {011004} (\bibinfo {year} {2019})}\BibitemShut {NoStop}%
\bibitem [{\citenamefont {Zhang}\ \emph {et~al.}(2017)\citenamefont {Zhang},
  \citenamefont {Pagano}, \citenamefont {Hess}, \citenamefont {Kyprianidis},
  \citenamefont {Becker}, \citenamefont {Kaplan}, \citenamefont {Gorshkov},
  \citenamefont {Gong},\ and\ \citenamefont {Monroe}}]{zhang2017observation}%
  \BibitemOpen
  \bibfield  {author} {\bibinfo {author} {\bibfnamefont {J.}~\bibnamefont
  {Zhang}}, \bibinfo {author} {\bibfnamefont {G.}~\bibnamefont {Pagano}},
  \bibinfo {author} {\bibfnamefont {P.~W.}\ \bibnamefont {Hess}}, \bibinfo
  {author} {\bibfnamefont {A.}~\bibnamefont {Kyprianidis}}, \bibinfo {author}
  {\bibfnamefont {P.}~\bibnamefont {Becker}}, \bibinfo {author} {\bibfnamefont
  {H.}~\bibnamefont {Kaplan}}, \bibinfo {author} {\bibfnamefont {A.~V.}\
  \bibnamefont {Gorshkov}}, \bibinfo {author} {\bibfnamefont {Z.-X.}\
  \bibnamefont {Gong}},\ and\ \bibinfo {author} {\bibfnamefont
  {C.}~\bibnamefont {Monroe}},\ }\bibfield  {title} {\bibinfo {title}
  {Observation of a many-body dynamical phase transition with a 53-qubit
  quantum simulator},\ }\href@noop {} {\bibfield  {journal} {\bibinfo
  {journal} {Nature}\ }\textbf {\bibinfo {volume} {551}},\ \bibinfo {pages}
  {601} (\bibinfo {year} {2017})}\BibitemShut {NoStop}%
\bibitem [{\citenamefont {Kaplan}\ \emph {et~al.}(2020)\citenamefont {Kaplan},
  \citenamefont {Guo}, \citenamefont {Tan}, \citenamefont {De}, \citenamefont
  {Marquardt}, \citenamefont {Pagano},\ and\ \citenamefont
  {Monroe}}]{PhysRevLett.125.120605}%
  \BibitemOpen
  \bibfield  {author} {\bibinfo {author} {\bibfnamefont {H.~B.}\ \bibnamefont
  {Kaplan}}, \bibinfo {author} {\bibfnamefont {L.}~\bibnamefont {Guo}},
  \bibinfo {author} {\bibfnamefont {W.~L.}\ \bibnamefont {Tan}}, \bibinfo
  {author} {\bibfnamefont {A.}~\bibnamefont {De}}, \bibinfo {author}
  {\bibfnamefont {F.}~\bibnamefont {Marquardt}}, \bibinfo {author}
  {\bibfnamefont {G.}~\bibnamefont {Pagano}},\ and\ \bibinfo {author}
  {\bibfnamefont {C.}~\bibnamefont {Monroe}},\ }\bibfield  {title} {\bibinfo
  {title} {Many-body dephasing in a trapped-ion quantum simulator},\ }\href
  {https://doi.org/10.1103/PhysRevLett.125.120605} {\bibfield  {journal}
  {\bibinfo  {journal} {Phys. Rev. Lett.}\ }\textbf {\bibinfo {volume} {125}},\
  \bibinfo {pages} {120605} (\bibinfo {year} {2020})}\BibitemShut {NoStop}%
\bibitem [{\citenamefont {Choi}\ \emph {et~al.}(2016)\citenamefont {Choi},
  \citenamefont {Hild}, \citenamefont {Zeiher}, \citenamefont {Schau{\ss}},
  \citenamefont {Rubio-Abadal}, \citenamefont {Yefsah}, \citenamefont
  {Khemani}, \citenamefont {Huse}, \citenamefont {Bloch},\ and\ \citenamefont
  {Gross}}]{Choi1547}%
  \BibitemOpen
  \bibfield  {author} {\bibinfo {author} {\bibfnamefont {J.-y.}\ \bibnamefont
  {Choi}}, \bibinfo {author} {\bibfnamefont {S.}~\bibnamefont {Hild}}, \bibinfo
  {author} {\bibfnamefont {J.}~\bibnamefont {Zeiher}}, \bibinfo {author}
  {\bibfnamefont {P.}~\bibnamefont {Schau{\ss}}}, \bibinfo {author}
  {\bibfnamefont {A.}~\bibnamefont {Rubio-Abadal}}, \bibinfo {author}
  {\bibfnamefont {T.}~\bibnamefont {Yefsah}}, \bibinfo {author} {\bibfnamefont
  {V.}~\bibnamefont {Khemani}}, \bibinfo {author} {\bibfnamefont {D.~A.}\
  \bibnamefont {Huse}}, \bibinfo {author} {\bibfnamefont {I.}~\bibnamefont
  {Bloch}},\ and\ \bibinfo {author} {\bibfnamefont {C.}~\bibnamefont {Gross}},\
  }\bibfield  {title} {\bibinfo {title} {Exploring the many-body localization
  transition in two dimensions},\ }\href
  {https://doi.org/10.1126/science.aaf8834} {\bibfield  {journal} {\bibinfo
  {journal} {Science}\ }\textbf {\bibinfo {volume} {352}},\ \bibinfo {pages}
  {1547} (\bibinfo {year} {2016})}\BibitemShut {NoStop}%
\bibitem [{\citenamefont {Kaufman}\ \emph {et~al.}(2016)\citenamefont
  {Kaufman}, \citenamefont {Tai}, \citenamefont {Lukin}, \citenamefont
  {Rispoli}, \citenamefont {Schittko}, \citenamefont {Preiss},\ and\
  \citenamefont {Greiner}}]{Kaufman794}%
  \BibitemOpen
  \bibfield  {author} {\bibinfo {author} {\bibfnamefont {A.~M.}\ \bibnamefont
  {Kaufman}}, \bibinfo {author} {\bibfnamefont {M.~E.}\ \bibnamefont {Tai}},
  \bibinfo {author} {\bibfnamefont {A.}~\bibnamefont {Lukin}}, \bibinfo
  {author} {\bibfnamefont {M.}~\bibnamefont {Rispoli}}, \bibinfo {author}
  {\bibfnamefont {R.}~\bibnamefont {Schittko}}, \bibinfo {author}
  {\bibfnamefont {P.~M.}\ \bibnamefont {Preiss}},\ and\ \bibinfo {author}
  {\bibfnamefont {M.}~\bibnamefont {Greiner}},\ }\bibfield  {title} {\bibinfo
  {title} {Quantum thermalization through entanglement in an isolated many-body
  system},\ }\href {https://doi.org/10.1126/science.aaf6725} {\bibfield
  {journal} {\bibinfo  {journal} {Science}\ }\textbf {\bibinfo {volume}
  {353}},\ \bibinfo {pages} {794} (\bibinfo {year} {2016})}\BibitemShut
  {NoStop}%
\bibitem [{\citenamefont {Houck}\ \emph {et~al.}(2012)\citenamefont {Houck},
  \citenamefont {T{\"u}reci},\ and\ \citenamefont {Koch}}]{Houck2012}%
  \BibitemOpen
  \bibfield  {author} {\bibinfo {author} {\bibfnamefont {A.~A.}\ \bibnamefont
  {Houck}}, \bibinfo {author} {\bibfnamefont {H.~E.}\ \bibnamefont
  {T{\"u}reci}},\ and\ \bibinfo {author} {\bibfnamefont {J.}~\bibnamefont
  {Koch}},\ }\bibfield  {title} {\bibinfo {title} {On-chip quantum simulation
  with superconducting circuits},\ }\href {https://doi.org/10.1038/nphys2251}
  {\bibfield  {journal} {\bibinfo  {journal} {Nature Physics}\ }\textbf
  {\bibinfo {volume} {8}},\ \bibinfo {pages} {292} (\bibinfo {year}
  {2012})}\BibitemShut {NoStop}%
\bibitem [{\citenamefont {Xu}\ \emph {et~al.}(2018)\citenamefont {Xu},
  \citenamefont {Chen}, \citenamefont {Zeng}, \citenamefont {Zhang},
  \citenamefont {Song}, \citenamefont {Liu}, \citenamefont {Guo}, \citenamefont
  {Zhang}, \citenamefont {Xu}, \citenamefont {Deng}, \citenamefont {Huang},
  \citenamefont {Wang}, \citenamefont {Zhu}, \citenamefont {Zheng},\ and\
  \citenamefont {Fan}}]{PhysRevLett.120.050507}%
  \BibitemOpen
  \bibfield  {author} {\bibinfo {author} {\bibfnamefont {K.}~\bibnamefont
  {Xu}}, \bibinfo {author} {\bibfnamefont {J.-J.}\ \bibnamefont {Chen}},
  \bibinfo {author} {\bibfnamefont {Y.}~\bibnamefont {Zeng}}, \bibinfo {author}
  {\bibfnamefont {Y.-R.}\ \bibnamefont {Zhang}}, \bibinfo {author}
  {\bibfnamefont {C.}~\bibnamefont {Song}}, \bibinfo {author} {\bibfnamefont
  {W.}~\bibnamefont {Liu}}, \bibinfo {author} {\bibfnamefont {Q.}~\bibnamefont
  {Guo}}, \bibinfo {author} {\bibfnamefont {P.}~\bibnamefont {Zhang}}, \bibinfo
  {author} {\bibfnamefont {D.}~\bibnamefont {Xu}}, \bibinfo {author}
  {\bibfnamefont {H.}~\bibnamefont {Deng}}, \bibinfo {author} {\bibfnamefont
  {K.}~\bibnamefont {Huang}}, \bibinfo {author} {\bibfnamefont
  {H.}~\bibnamefont {Wang}}, \bibinfo {author} {\bibfnamefont {X.}~\bibnamefont
  {Zhu}}, \bibinfo {author} {\bibfnamefont {D.}~\bibnamefont {Zheng}},\ and\
  \bibinfo {author} {\bibfnamefont {H.}~\bibnamefont {Fan}},\ }\bibfield
  {title} {\bibinfo {title} {Emulating many-body localization with a
  superconducting quantum processor},\ }\href
  {https://doi.org/10.1103/PhysRevLett.120.050507} {\bibfield  {journal}
  {\bibinfo  {journal} {Phys. Rev. Lett.}\ }\textbf {\bibinfo {volume} {120}},\
  \bibinfo {pages} {050507} (\bibinfo {year} {2018})}\BibitemShut {NoStop}%
\bibitem [{\citenamefont {Pierangeli}\ \emph {et~al.}(2019)\citenamefont
  {Pierangeli}, \citenamefont {Marcucci},\ and\ \citenamefont
  {Conti}}]{pierangeli2019large}%
  \BibitemOpen
  \bibfield  {author} {\bibinfo {author} {\bibfnamefont {D.}~\bibnamefont
  {Pierangeli}}, \bibinfo {author} {\bibfnamefont {G.}~\bibnamefont
  {Marcucci}},\ and\ \bibinfo {author} {\bibfnamefont {C.}~\bibnamefont
  {Conti}},\ }\bibfield  {title} {\bibinfo {title} {Large-scale photonic ising
  machine by spatial light modulation},\ }\href@noop {} {\bibfield  {journal}
  {\bibinfo  {journal} {Physical review letters}\ }\textbf {\bibinfo {volume}
  {122}},\ \bibinfo {pages} {213902} (\bibinfo {year} {2019})}\BibitemShut
  {NoStop}%
\bibitem [{\citenamefont {Pierangeli}\ \emph {et~al.}(2021)\citenamefont
  {Pierangeli}, \citenamefont {Rafayelyan}, \citenamefont {Conti},\ and\
  \citenamefont {Gigan}}]{pierangeli2020scalable}%
  \BibitemOpen
  \bibfield  {author} {\bibinfo {author} {\bibfnamefont {D.}~\bibnamefont
  {Pierangeli}}, \bibinfo {author} {\bibfnamefont {M.}~\bibnamefont
  {Rafayelyan}}, \bibinfo {author} {\bibfnamefont {C.}~\bibnamefont {Conti}},\
  and\ \bibinfo {author} {\bibfnamefont {S.}~\bibnamefont {Gigan}},\ }\bibfield
   {title} {\bibinfo {title} {Scalable spin-glass optical simulator},\ }\href
  {https://doi.org/10.1103/PhysRevApplied.15.034087} {\bibfield  {journal}
  {\bibinfo  {journal} {Phys. Rev. Applied}\ }\textbf {\bibinfo {volume}
  {15}},\ \bibinfo {pages} {034087} (\bibinfo {year} {2021})}\BibitemShut
  {NoStop}%
\bibitem [{\citenamefont {Weinberg}\ and\ \citenamefont
  {Bukov}(2017)}]{weinberg2016quspin}%
  \BibitemOpen
  \bibfield  {author} {\bibinfo {author} {\bibfnamefont {P.}~\bibnamefont
  {Weinberg}}\ and\ \bibinfo {author} {\bibfnamefont {M.}~\bibnamefont
  {Bukov}},\ }\bibfield  {title} {\bibinfo {title} {{QuSpin: a Python Package
  for Dynamics and Exact Diagonalisation of Quantum Many Body Systems part I:
  spin chains}},\ }\href {https://doi.org/10.21468/SciPostPhys.2.1.003}
  {\bibfield  {journal} {\bibinfo  {journal} {SciPost Phys.}\ }\textbf
  {\bibinfo {volume} {2}},\ \bibinfo {pages} {003} (\bibinfo {year}
  {2017})}\BibitemShut {NoStop}%
\bibitem [{\citenamefont {Johansson}\ \emph {et~al.}(2012)\citenamefont
  {Johansson}, \citenamefont {Nation},\ and\ \citenamefont
  {Nori}}]{johansson2012qutip}%
  \BibitemOpen
  \bibfield  {author} {\bibinfo {author} {\bibfnamefont {J.~R.}\ \bibnamefont
  {Johansson}}, \bibinfo {author} {\bibfnamefont {P.~D.}\ \bibnamefont
  {Nation}},\ and\ \bibinfo {author} {\bibfnamefont {F.}~\bibnamefont {Nori}},\
  }\bibfield  {title} {\bibinfo {title} {Qutip: An open-source python framework
  for the dynamics of open quantum systems},\ }\href@noop {} {\bibfield
  {journal} {\bibinfo  {journal} {Computer Physics Communications}\ }\textbf
  {\bibinfo {volume} {183}},\ \bibinfo {pages} {1760} (\bibinfo {year}
  {2012})}\BibitemShut {NoStop}%
\bibitem [{\citenamefont {Ilievski}\ \emph {et~al.}(2015)\citenamefont
  {Ilievski}, \citenamefont {De~Nardis}, \citenamefont {Wouters}, \citenamefont
  {Caux}, \citenamefont {Essler},\ and\ \citenamefont
  {Prosen}}]{ilievski2015complete}%
  \BibitemOpen
  \bibfield  {author} {\bibinfo {author} {\bibfnamefont {E.}~\bibnamefont
  {Ilievski}}, \bibinfo {author} {\bibfnamefont {J.}~\bibnamefont {De~Nardis}},
  \bibinfo {author} {\bibfnamefont {B.}~\bibnamefont {Wouters}}, \bibinfo
  {author} {\bibfnamefont {J.-S.}\ \bibnamefont {Caux}}, \bibinfo {author}
  {\bibfnamefont {F.~H.}\ \bibnamefont {Essler}},\ and\ \bibinfo {author}
  {\bibfnamefont {T.}~\bibnamefont {Prosen}},\ }\bibfield  {title} {\bibinfo
  {title} {Complete generalized {Gibbs} ensembles in an interacting theory},\
  }\href@noop {} {\bibfield  {journal} {\bibinfo  {journal} {Physical Review
  Letters}\ }\textbf {\bibinfo {volume} {115}},\ \bibinfo {pages} {157201}
  (\bibinfo {year} {2015})}\BibitemShut {NoStop}%
\bibitem [{\citenamefont {Vosk}\ and\ \citenamefont
  {Altman}(2014)}]{vosk2014dynamical}%
  \BibitemOpen
  \bibfield  {author} {\bibinfo {author} {\bibfnamefont {R.}~\bibnamefont
  {Vosk}}\ and\ \bibinfo {author} {\bibfnamefont {E.}~\bibnamefont {Altman}},\
  }\bibfield  {title} {\bibinfo {title} {Dynamical quantum phase transitions in
  random spin chains},\ }\href@noop {} {\bibfield  {journal} {\bibinfo
  {journal} {Physical Review Letters}\ }\textbf {\bibinfo {volume} {112}},\
  \bibinfo {pages} {217204} (\bibinfo {year} {2014})}\BibitemShut {NoStop}%
\bibitem [{\citenamefont {Geraedts}\ \emph {et~al.}(2017)\citenamefont
  {Geraedts}, \citenamefont {Bhatt},\ and\ \citenamefont
  {Nandkishore}}]{geraedts2017emergent}%
  \BibitemOpen
  \bibfield  {author} {\bibinfo {author} {\bibfnamefont {S.~D.}\ \bibnamefont
  {Geraedts}}, \bibinfo {author} {\bibfnamefont {R.~N.}\ \bibnamefont
  {Bhatt}},\ and\ \bibinfo {author} {\bibfnamefont {R.}~\bibnamefont
  {Nandkishore}},\ }\bibfield  {title} {\bibinfo {title} {Emergent local
  integrals of motion without a complete set of localized eigenstates},\
  }\href@noop {} {\bibfield  {journal} {\bibinfo  {journal} {Physical Review
  B}\ }\textbf {\bibinfo {volume} {95}},\ \bibinfo {pages} {064204} (\bibinfo
  {year} {2017})}\BibitemShut {NoStop}%
\bibitem [{\citenamefont {Zobov}\ and\ \citenamefont
  {Lundin}(2017)}]{zobov2017effect}%
  \BibitemOpen
  \bibfield  {author} {\bibinfo {author} {\bibfnamefont {V.~E.}\ \bibnamefont
  {Zobov}}\ and\ \bibinfo {author} {\bibfnamefont {A.~A.}\ \bibnamefont
  {Lundin}},\ }\bibfield  {title} {\bibinfo {title} {On the effect of an
  inhomogeneous magnetic field and many-body localization on the increase in
  the second moment of multiple-quantum nmr with time},\ }\href@noop {}
  {\bibfield  {journal} {\bibinfo  {journal} {JETP Letters}\ }\textbf {\bibinfo
  {volume} {105}},\ \bibinfo {pages} {514} (\bibinfo {year}
  {2017})}\BibitemShut {NoStop}%
\bibitem [{\citenamefont {Wei}\ \emph {et~al.}(2018)\citenamefont {Wei},
  \citenamefont {Ramanathan},\ and\ \citenamefont
  {Cappellaro}}]{wei2018exploring}%
  \BibitemOpen
  \bibfield  {author} {\bibinfo {author} {\bibfnamefont {K.~X.}\ \bibnamefont
  {Wei}}, \bibinfo {author} {\bibfnamefont {C.}~\bibnamefont {Ramanathan}},\
  and\ \bibinfo {author} {\bibfnamefont {P.}~\bibnamefont {Cappellaro}},\
  }\bibfield  {title} {\bibinfo {title} {Exploring localization in nuclear spin
  chains},\ }\href@noop {} {\bibfield  {journal} {\bibinfo  {journal} {Physical
  Review Letters}\ }\textbf {\bibinfo {volume} {120}},\ \bibinfo {pages}
  {070501} (\bibinfo {year} {2018})}\BibitemShut {NoStop}%
\bibitem [{\citenamefont {Richerme}\ \emph {et~al.}(2014)\citenamefont
  {Richerme}, \citenamefont {Gong}, \citenamefont {Lee}, \citenamefont {Senko},
  \citenamefont {Smith}, \citenamefont {Foss-Feig}, \citenamefont {Michalakis},
  \citenamefont {Gorshkov},\ and\ \citenamefont {Monroe}}]{richerme2014non}%
  \BibitemOpen
  \bibfield  {author} {\bibinfo {author} {\bibfnamefont {P.}~\bibnamefont
  {Richerme}}, \bibinfo {author} {\bibfnamefont {Z.-X.}\ \bibnamefont {Gong}},
  \bibinfo {author} {\bibfnamefont {A.}~\bibnamefont {Lee}}, \bibinfo {author}
  {\bibfnamefont {C.}~\bibnamefont {Senko}}, \bibinfo {author} {\bibfnamefont
  {J.}~\bibnamefont {Smith}}, \bibinfo {author} {\bibfnamefont
  {M.}~\bibnamefont {Foss-Feig}}, \bibinfo {author} {\bibfnamefont
  {S.}~\bibnamefont {Michalakis}}, \bibinfo {author} {\bibfnamefont {A.~V.}\
  \bibnamefont {Gorshkov}},\ and\ \bibinfo {author} {\bibfnamefont
  {C.}~\bibnamefont {Monroe}},\ }\bibfield  {title} {\bibinfo {title}
  {Non-local propagation of correlations in quantum systems with long-range
  interactions},\ }\href@noop {} {\bibfield  {journal} {\bibinfo  {journal}
  {Nature}\ }\textbf {\bibinfo {volume} {511}},\ \bibinfo {pages} {198}
  (\bibinfo {year} {2014})}\BibitemShut {NoStop}%
\bibitem [{\citenamefont {Huang}\ \emph {et~al.}(2020)\citenamefont {Huang},
  \citenamefont {Kueng},\ and\ \citenamefont {Preskill}}]{huang2020predicting}%
  \BibitemOpen
  \bibfield  {author} {\bibinfo {author} {\bibfnamefont {H.-Y.}\ \bibnamefont
  {Huang}}, \bibinfo {author} {\bibfnamefont {R.}~\bibnamefont {Kueng}},\ and\
  \bibinfo {author} {\bibfnamefont {J.}~\bibnamefont {Preskill}},\ }\bibfield
  {title} {\bibinfo {title} {Predicting many properties of a quantum system
  from very few measurements},\ }\href@noop {} {\bibfield  {journal} {\bibinfo
  {journal} {Nature Physics}\ }\textbf {\bibinfo {volume} {16}},\ \bibinfo
  {pages} {1050} (\bibinfo {year} {2020})}\BibitemShut {NoStop}%
\bibitem [{\citenamefont {Kubota}\ \emph {et~al.}(2019)\citenamefont {Kubota},
  \citenamefont {Nakajima},\ and\ \citenamefont
  {Takahashi}}]{kubota2019dynamical}%
  \BibitemOpen
  \bibfield  {author} {\bibinfo {author} {\bibfnamefont {T.}~\bibnamefont
  {Kubota}}, \bibinfo {author} {\bibfnamefont {K.}~\bibnamefont {Nakajima}},\
  and\ \bibinfo {author} {\bibfnamefont {H.}~\bibnamefont {Takahashi}},\
  }\bibfield  {title} {\bibinfo {title} {Dynamical anatomy of narma10 benchmark
  task},\ }\href@noop {} {\bibfield  {journal} {\bibinfo  {journal} {arXiv
  preprint arXiv:1906.04608}\ } (\bibinfo {year} {2019})}\BibitemShut {NoStop}%
\end{thebibliography}%


\end{document}